\documentclass[preprint,12pt,authoryear]{elsarticle}

\usepackage{amssymb}
\usepackage{amsmath}
\usepackage{lmodern}

\usepackage{booktabs}

\makeatletter
\newcounter{algorithm}
\newenvironment{algorithm}{%
  \par\addvspace{\medskipamount}\noindent
  \begin{minipage}{\linewidth}\def\@captype{algorithm}%
}{%
  \end{minipage}\par\addvspace{\medskipamount}%
}
\newcommand{\ext@algorithm}{loa}
\newcommand{\fnum@algorithm}{Algorithm~\thealgorithm}
\makeatother

\usepackage{bm}
\usepackage{xcolor,tikz}
\usetikzlibrary{arrows.meta,calc,fit}

\colorlet{revision}{black}
\newcommand{\rev}[1]{{\color{revision}#1}}

\usepackage{xurl}
\usepackage[hidelinks]{hyperref}
\providecommand{\doi}[1]{\href{https://doi.org/#1}{\nolinkurl{#1}}}

\graphicspath{{figures/}}

\begin{document}

\hypersetup{
  pdftitle={Direct Numerical Simulations of Incompressible Homogeneous Turbulence under General Spatially Uniform Mean Velocity Gradients},
  pdfauthor={Tamerlan Hajiyev, Emre Sefer, Ivan Otic, Ismail Ar{\i}, {\"O}zg{\"u}r Ertun\c{c}}
}

\begin{frontmatter}



\title{Direct Numerical \rev{Simulations} of Incompressible Homogeneous Turbulence under General Spatially Uniform Mean Velocity Gradients}


\author[cs]{Tamerlan Hajiyev}
\ead{tamerlan.hajiyev@ozu.edu.tr}
\author[ai]{Emre Sefer}
\ead{emre.sefer@ozyegin.edu.tr}
\author[kit]{Ivan Otic}
\ead{ivan.otic@kit.edu}
\author[cs]{Ismail Ar{\i}}
\ead{ismail.ari@ozyegin.edu.tr}
\author[me]{{\"O}zg{\"u}r Ertun\c{c}\corref{cor1}}
\ead{ozgur.ertunc@ozyegin.edu.tr}
\cortext[cor1]{Corresponding author}

\affiliation[cs]{organization={Department of Computer Science, {\"O}zye{\u{g}}in University},
            city={Istanbul},
            country={T\"urkiye}}
\affiliation[ai]{organization={Department of Artificial Intelligence and Data Engineering, \\ {\"O}zye{\u{g}}in University},
            city={Istanbul},
            country={T\"urkiye}}
\affiliation[kit]{organization={Karlsruhe Institute of Technology},
            city={Karlsruhe},
            country={Germany}}
\affiliation[me]{organization={Fluid Dynamics and Spray Laboratory, Department of Mechanical Engineering, \\{\"O}zye{\u{g}}in University},
            city={Istanbul},
            country={T\"urkiye}}

\begin{abstract}
We extend \citeauthor{rogallo1981}'s (\citeyear{rogallo1981}) pseudospectral
algorithm for direct numerical simulation (DNS) of incompressible homogeneous
turbulence \rev{distorted by} general spatially uniform mean velocity gradients. We
generalize the velocity transformation to a full matrix form and use fourth-order
commutator-free Magnus integration to advance the transformation matrices,
preserving their unit determinants by construction. As deformation accumulates,
physical resolution can deteriorate in some directions. We address this with a
general three-dimensional remeshing algorithm that extends the method beyond
simple shear. To represent the same physical lattice in a new basis, a bounded
search generates candidate bases using integer matrices with determinant one,
constructed through compositions of elementary column operations.
We rank the eligible candidate bases by physical spectral coverage and basis
conditioning, then select the best among those whose predicted kinetic energy
and dissipation losses satisfy specified tolerances. Our solver agrees with
established homogeneous turbulence benchmarks \rev{from
\citeauthor{leeReynolds1985} (\citeyear{leeReynolds1985}) and
\citeauthor{blaisdellShariff1996} (\citeyear{blaisdellShariff1996})}.
\rev{Simulations with and without remeshing were compared across several mean
deformations. In the general shear cases with mean rotation and strain, the
unremeshed box lost physical resolution in some directions, and the runs
differed by up to $24\%$ in shear stress and $66\%$ in the dissipation tensor
component $\epsilon_{23}$, highlighting the importance of remeshing.}
\end{abstract}

\begin{graphicalabstract}
\par
\centering
\setlength{\linewidth}{\hsize}
\noindent\includegraphics[width=\linewidth]{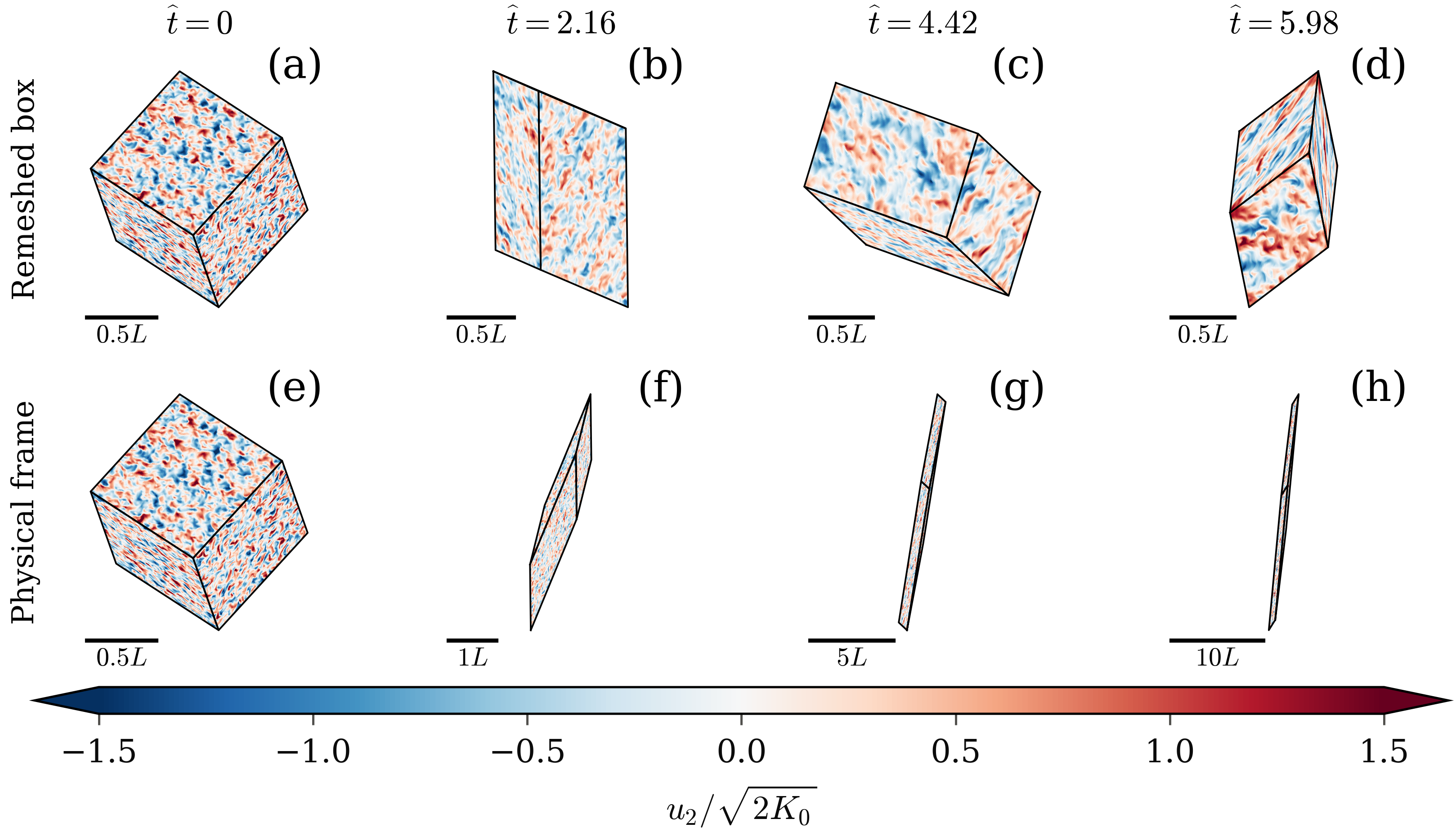}\par
\medskip
{\small\raggedright\noindent
\rev{Our solver generalizes the remeshing algorithm beyond simple shear, making it possible
to simulate a broader range of homogeneous anisotropic turbulent flows.
Under three-dimensional shear with mean strain and rotation,
the original box becomes highly elongated (bottom row). Remeshing gives the solver
a less distorted box for the same flow (top row). Colours show the
velocity fluctuation along $x_2$, normalized by the initial Root Mean Square (RMS)
fluctuation speed. Each panel is scaled to fit, and all scale bars give physical
distances in units of the same initial box side $L$.}\par}
\end{graphicalabstract}

\begin{highlights}
\item \rev{Our solver makes it possible to simulate homogeneous anisotropic turbulence distorted by general
spatially uniform mean deformations.}
\item \rev{\citeauthor{rogallo1981}'s (\citeyear{rogallo1981}) shear
remeshing is a special case of our general remeshing algorithm.}
\item \rev{A full matrix velocity map generalizes \citeauthor{rogallo1981}'s
(\citeyear{rogallo1981}) rapid distortion variables, and time integration
preserves its unit determinant.}
\item \rev{Without remeshing, loss of resolution can lead to differences of up to
66\% in direction-dependent turbulence statistics.}
\item \rev{The solver agrees with established benchmarks for canonical flows.}
\end{highlights}


\end{frontmatter}


\section{Introduction}
\label{sec:intro}

Flows under spatially uniform mean velocity gradients provide a controlled setting for
studying the response of homogeneous turbulence to strain and rotation. The mean
velocity varies linearly in space, while the turbulent fluctuations remain statistically
homogeneous. This allows the response to mean deformation to be studied without walls
or spatial inhomogeneity in the turbulence. \rev{Pioneering direct numerical
simulation studies by \citet{rogallo1981} and \citet{leeReynolds1985} examined
the evolution and structure of homogeneous turbulence under plane strain,
axisymmetric contraction and expansion, and simple shear.}

Direct numerical simulations \rev{(DNS)} provide reference data for turbulence modelling
\citep{pope2000,duraisamy2019}, including the structural information needed by
closures for homogeneous anisotropic turbulence \citep{kassinosReynoldsRogers2001}.
\rev{Existing numerical and experimental studies have considered irrotational
plane strain \citep{townsend1954}, axisymmetric contraction, expansion, and
successive straining \citep{ertunc2007},} successive plane strains with changing
principal directions \citep{genceMathieu1979}, elliptic flows and homogeneous
shear \citep{blaisdellShariff1996}, periodic shear \citep{yuGirimaji2006},
straining and destraining cycles \citep{gualtieriMeneveau2010}, and
time-dependent axisymmetric contraction followed by relaxation
\citep{clayYeung2016}.
\rev{These canonical cases provide valuable reference data for testing turbulence
closures. \citet*{jovanovic2003} used DNS and experimental data to assess an
anisotropy-invariant closure for axisymmetric turbulence. Later studies compared
Reynolds stress closures against measurements under axisymmetric strain
\citep{ertuncEtAl2007,banerjeeEtAl2009PressureStrain}. However, assessing closures under
more general conditions requires flows with broader combinations of strain and
rotation.}
High-fidelity data for such conditions remain limited \citep{dongHaeringMoser2025}.
Such data are also needed to extend our framework for constructing deployable
hard-constrained turbulence closures of unclosed terms in the transport
equations of Reynolds stresses and turbulent dissipation rate
\citep{hajiyev2026closure} to anisotropic flows.

The deforming coordinate method of \citet{rogallo1981} provides a foundation for
pseudospectral DNS under mean deformation. Its general equations admit time-dependent
mean velocity gradients, although the reported implementation uses a restricted form.
\citet{leeReynolds1985} applied Rogallo's approach to homogeneous turbulence
under plane strain, axisymmetric contraction and axisymmetric expansion,
and \citet{blaisdellShariff1996} modified his code to simulate elliptic
streamline flows. For their elliptic cases, they avoided remeshing by restricting
the streamline aspect ratios so that distortion of the computational box remained
limited.

The main difficulty is distortion of the physical periodic box, which reduces resolution in
some directions. \citet{brucker2007} developed a method that avoids remeshing for
homogeneous shear, while remeshing under arbitrary mean velocity gradients remains a difficulty
\citep{zhu2019thesis}. For molecular simulations of steady three-dimensional flows,
\citet{dobson2014} \rev{provides a remeshing algorithm using} integer powers of two
prescribed commuting matrices to bound the distortion of the periodic box
under constant mean velocity gradients. However, his algorithm uses fixed
stretch rates for choosing the matrix powers, and is limited to steady mean flows.

The second difficulty is in preserving volume. Under incompressible mean flow, an initial
unit determinant of the deformation mapping must remain unity, but numerical integration
does not automatically enforce this constraint. \citet{girimajiPope1990} examined this
requirement using fourth-order integration. To preserve this constraint, Lie group
methods can construct updates from exponentials of trace-free matrices
\citep{iserles2000}.

To address these difficulties, we extend the algorithm of \citet{rogallo1981}
for incompressible homogeneous turbulence to flows under general spatially uniform
mean velocity gradients. Within this formulation, a full matrix velocity
transformation
absorbs the direct mean velocity gradient term in the fluctuation equations,
while fourth-order commutator-free Magnus updates preserve unit determinants.
These updates use the established two-exponential integrator of
\citet{blanesMoan2006} for coordinate and velocity transformations. As
deformation accumulates, a three-dimensional remeshing procedure searches over
integer changes of lattice basis to select an equivalent basis of the evolving
physical periodic domain. Candidate bases are accepted only if their predicted
kinetic energy and dissipation losses\rev{, resulting from Fourier modes discarded
during remeshing,} satisfy specified tolerances. We then assess the method against
published homogeneous turbulence benchmarks \rev{\citep{leeReynolds1985,blaisdellShariff1996}}
and demonstrate its applicability to a wider range of mean deformations.

\section{Governing equations and transformations}
\label{sec:governing}

We consider an incompressible fluid with constant kinematic viscosity $\nu$.
\rev{We start from the Navier--Stokes equations in physical coordinates
\citep{rogallo1981} and include a body force per unit mass $F_i$.}

{\color{revision}
\begin{equation}
\begin{aligned}
 \frac{\partial u_i^{\rm tot}}{\partial t}
 +u_j^{\rm tot}\frac{\partial u_i^{\rm tot}}{\partial x_j}
 &=-\frac{1}{\rho}\frac{\partial P^{\rm tot}}{\partial x_i}
   +\nu\nabla^2 u_i^{\rm tot}+F_i,\\
 \frac{\partial u_i^{\rm tot}}{\partial x_i}&=0.
\end{aligned}
 \label{eq:total_navier_stokes}
\end{equation}
}

\rev{Here $P^{\rm tot}$ is the total pressure and $\rho$ is the constant density.
We write $P^{\rm tot}=\overline P+p$, where $\overline P$ and $p$ are the mean
and fluctuating pressures, respectively. We decompose the total velocity as}
$u_i^{\rm tot}=A_{ij}(t)x_j+u_i$, where
$A_{ij}=\partial U_i/\partial x_j$ is the mean velocity gradient.
\rev{Following \citet{rogallo1981}, we prescribe a spatially uniform mean velocity
gradient with $\partial U_i/\partial x_i=A_{ii}=0$ and ensemble average the
equations in coordinates moving with the mean flow. Under the ergodic hypothesis,
we estimate ensemble averages $\langle\cdot\rangle$ by volume averaging the
homogeneous fluctuations, with $\langle u_i\rangle=0$. Homogeneity removes the
Reynolds stress divergence, and the mean viscous term $\nu\nabla^2 U_i$ vanishes
because $U_i$ is linear in space.}

\rev{To support the prescribed mean deformation \citep{yuGirimaji2007}, we include
a deterministic body force per unit mass of the form}

{\color{revision}
\begin{equation}
 F_i=G_{ij}(t)x_j.
 \label{eq:linear_body_force}
\end{equation}
}

The mean-momentum equation requires the pressure gradient
$\partial\overline P/\partial x_i=P_{ij}x_j$
\rev{\citep{rogallo1981,hajiyev2026rogallo}. The resulting balance is}

\begin{equation}
 \dot{\mathbf A}+\mathbf A^2+\rev{\frac{\mathbf P}{\rho}}=\mathbf G,
 \qquad P_{ij}=\frac{\partial^2\overline P}{\partial x_i\partial x_j}.
 \label{eq:mean_momentum}
\end{equation}
To separate strain and rotation, we write
$\mathbf A=\mathbf S+\boldsymbol{\Omega}$, where

\begin{equation}
 \mathbf S=\frac12(\mathbf A+\mathbf A^T),
 \qquad
 \boldsymbol{\Omega}=\frac12(\mathbf A-\mathbf A^T).
 \label{eq:mean_gradient_parts}
\end{equation}

We decompose $\mathbf G=\mathbf G_S+\mathbf G_{\Omega}$, with
$\mathbf G_S=(\mathbf G+\mathbf G^{\mathsf T})/2$ and
$\mathbf G_{\Omega}=(\mathbf G-\mathbf G^{\mathsf T})/2$.
Taking the symmetric and antisymmetric parts of the mean momentum
equation then gives

\begin{equation}
 \dot{\mathbf S}+\mathbf S^2+\boldsymbol{\Omega}^2+\rev{\frac{\mathbf P}{\rho}}
 =\mathbf G_S,
 \label{eq:mean_strain_balance}
\end{equation}
and
\begin{equation}
 \dot{\boldsymbol{\Omega}}
 +\mathbf S\boldsymbol{\Omega}+\boldsymbol{\Omega}\mathbf S
 =\mathbf G_{\Omega}.
 \label{eq:mean_rotation_balance}
\end{equation}

The symmetric linear forcing is conservative and may be absorbed into
the mean pressure, giving

\begin{equation}
\begin{aligned}
 \rev{\frac{\mathbf P}{\rho}}&=-\dot{\mathbf S}-\mathbf S^2-\boldsymbol{\Omega}^2,\\
 \mathbf G=\mathbf G_{\Omega}
 &=\dot{\boldsymbol{\Omega}}
   +\mathbf S\boldsymbol{\Omega}+\boldsymbol{\Omega}\mathbf S.
\end{aligned}
 \label{eq:prescribed_mean_support}
\end{equation}

The case $\mathbf G=0$ recovers the unforced mean flow evolution.
In the prescribed formulation, $\mathbf A(t)$ is supplied directly to
the solver and $\boldsymbol{\Omega}$ is not evolved separately. The mean
force has no fluctuating component and therefore does not appear in the
fluctuation equations. Detailed derivations are given
in \citet{hajiyev2026rogallo}.

In deforming coordinates
$\mathbf x'=\mathbf B(t)\mathbf x$, the fluctuation equations are

\begin{equation}
\begin{aligned}
  \frac{\partial u_i}{\partial t}+A_{ij}u_j
  +B_{kj}\frac{\partial(u_i u_j)}{\partial x_k'}
  &=-\rev{\frac{B_{ki}}{\rho}}\frac{\partial p}{\partial x_k'}
   +\nu B_{kj}B_{\ell j}
     \frac{\partial^2u_i}{\partial x_k'\partial x_\ell'},\\
  B_{ki}\frac{\partial u_i}{\partial x_k'}&=0.
\end{aligned}
  \label{eq:fluctuation_momentum}
\end{equation}

The coordinate transformation leaves the velocity components unchanged and
the time derivatives are taken at fixed $\mathbf x'$.

We evaluate kinetic energy, viscous dissipation for homogeneous turbulence,
and mean flow production \rev{following \citet{ertunc2025transport}}.
\begin{equation}
K=\frac12\langle u_i u_i\rangle,\qquad
\epsilon=\nu\left\langle
\left(\frac{\partial u_i}{\partial x_j}\right)^2
\right\rangle,\qquad
P=-A_{ij}\langle u_i u_j\rangle.
\label{eq:bulk_definitions}
\end{equation}
The derivatives in Eq.~\eqref{eq:bulk_definitions} are taken in
physical coordinates.

\subsection{Coordinate and velocity transformations}
\label{sec:deforming_coordinates}

The coordinate map $\mathbf x'=\mathbf B\mathbf x$ follows the mean
deformation and removes mean advection from the fluctuation equations,
while velocity fluctuations advect fluid through the fixed computational
grid with velocity $\mathbf B\mathbf u$.

\rev{Some global advancement variables can be expressed in an equivalent
step-local form \citep{hajiyev2026rogallo}. Since the velocity at $t_n$ already
carries the earlier evolution, these variables need only be evaluated from
$t_n$ to $t$, limiting their growth and the associated numerical error.}

\rev{For the restricted form of the mean velocity gradient, \citet{rogallo1981}
used the diagonal entries of $\mathbf B$ to absorb the diagonal deformation
terms that can make time integration stiff in the rapid distortion limit.
We generalize this scaling to the full matrix transformation}

{\color{revision}
\begin{equation}
  \mathbf u^*=\mathbf C_n^{-1}\mathbf u.
  \label{eq:velocity_transformation}
\end{equation}
}

\rev{The step-local velocity map $\mathbf C_n$ absorbs the full term
$\mathbf A\mathbf u$ in Eq.~\eqref{eq:fluctuation_momentum}, including coupling
between velocity components, and removes the stiffness associated with
advancing that term explicitly.} The maps satisfy

\begin{equation}
  \dot{\mathbf B}=-\mathbf B\mathbf A,
  \qquad
  \dot{\mathbf C}_n=-\mathbf A\mathbf C_n,
  \qquad \mathbf C_n(t_n)=\mathbf I.
  \label{eq:map_evolution}
\end{equation}

The multiplication orders differ because $\mathbf B$ transforms
coordinates whereas $\mathbf C_n$ transforms velocity. Starting from
$\mathbf B(t_0)=\mathbf I$, the coordinate map evolves between remeshes.
The velocity map is restarted at each timestep with $\mathbf C_n(t_n)=\mathbf I$;
its earlier evolution is already contained in the physical velocity carried to
the next step.
The relation between the step-local and global transformations is
derived in \citet{hajiyev2026rogallo}.

We now write the transformed velocity in Fourier space
\rev{\citep{rogallo1981,hajiyev2026rogallo}}. A tilde denotes
a coefficient with phase $\exp(i\mathbf k\cdot\mathbf x')$. The physical
wavevector is $\boldsymbol\kappa=\mathbf B^{\mathsf T}\mathbf k$, with
$\kappa^2=\boldsymbol\kappa\cdot\boldsymbol\kappa$. We also use
the vectors $\mathbf q=\mathbf C_n^{-1}\boldsymbol\kappa$ and
$\boldsymbol\tau=\mathbf C_n^{\mathsf T}\boldsymbol\kappa$.
After the velocity transformation, the equation for each Fourier mode contains
the viscous decay term $-\nu\kappa^2\widetilde u_i^*$, whose rate increases with
wavenumber. We remove this stiffness using $F_n$, then introduce the pressure
impulse $\phi$ to absorb the pressure term into the double-star advancement
variable,

\begin{equation}
\begin{aligned}
  F_n(\mathbf k,t)&=\exp\!\left[
    \nu\int_{t_n}^{t}\kappa^2(\mathbf k,s)\,ds\right],\\
  \dot\phi&=F_n\widetilde p,\\
  \widetilde u_i^{**}&=F_n\widetilde u_i^*+\rev{\frac{i q_i\phi}{\rho}}.
\end{aligned}
  \label{eq:advancement_variables}
\end{equation}

Because $\mathbf q$ changes under a general mean deformation, the
pressure impulse leaves a correction in the double-star equation,

\begin{equation}
  \frac{d\widetilde u_i^{**}}{dt}
  =-iF_n\tau_j\widetilde{u_i^*u_j^*}
  +\rev{\frac{i}{\rho}}(C_n^{-1})_{ij}(A_{j\ell}-A_{\ell j})\kappa_\ell\phi.
  \label{eq:advancement_equations}
\end{equation}

At each evaluation, continuity determines $\phi$ from the double-star
coefficients. We then recover the transformed and physical velocities,

\begin{equation}
  \phi=-\frac{i\rev{\rho}\tau_j\widetilde u_j^{**}}{\kappa^2},
  \qquad
  \widetilde u_i^*=\frac{\widetilde u_i^{**}-i q_i\phi\rev{/\rho}}{F_n},
  \qquad
  \widetilde u_i=(C_n)_{ij}\widetilde u_j^*.
  \label{eq:velocity_reconstruction}
\end{equation}

These expressions apply to nonzero wavevectors. The zero mode of the
fluctuating velocity is zero, and we set $\phi=0$ there. Nonlinear
products use the recovered single-star velocity. At the start of each
timestep, $\mathbf C_n=\mathbf I$ and $F_n=1$, so
$\widetilde{\mathbf u}^{**}(t_n)=\widetilde{\mathbf u}(t_n)$. The physical
velocity recovered at the endpoint initializes the next step. Detailed
derivations of the transformed equations and velocity reconstruction
are given in \citet{hajiyev2026rogallo}.

\subsection{\texorpdfstring{\rev{Rapid distortion limit}}{Rapid distortion limit}}

\rev{Following \citet{ertunc2007}, we define the dimensionless strain rate}

{\color{revision}
\begin{equation}
  S^*=\frac{2KS_{\mathrm{eq}}}{\epsilon},
  \qquad S_{\mathrm{eq}}=\left(\frac{S_{ij}S_{ij}}{2}\right)^{1/2},
  \qquad S_{ij}=\frac{A_{ij}+A_{ji}}{2}.
  \label{eq:rapid_distortion_parameter}
\end{equation}
}

\rev{For $S^*\gg1$, the mean strain time $1/S_{\mathrm{eq}}$ is much shorter
than the eddy turnover time $2K/\epsilon$. In the rapid distortion limit,
nonlinear interactions have a negligible effect on the velocity fluctuations
over the deformation interval \citep{batchelorProudman1954}. In viscous rapid
distortion theory (RDT), we retain the viscous factor $F_n$ and omit the
nonlinear term in Eq.~\eqref{eq:advancement_equations} to obtain}

{\color{revision}
\begin{equation}
  \frac{d\widetilde u_i^{**}}{dt}
  =\frac{i}{\rho}(C_n^{-1})_{ij}(A_{j\ell}-A_{\ell j})\kappa_\ell\phi.
  \label{eq:viscous_rdt}
\end{equation}
}

\subsection{Nondimensionalization}
\label{sec:nondimensionalization}

Following \citet{pope2000}, we define
reference scales $U_r=\sqrt{K_0}$, $L_r=K_0^{3/2}/\epsilon_0$, and
$T_r=L_r/U_r=K_0/\epsilon_0$, based on the initial field at $t_0=0$.
These scales remain fixed throughout each run. The dimensionless variables are
\begin{equation}
\begin{gathered}
 \widehat{\mathbf x}=\mathbf x/L_r,\quad
 \widehat t=t/T_r,\quad
 \widehat{\mathbf u}=\mathbf u/U_r,\quad
 \widehat p=p/(\rev{\rho}U_r^2),\\
 \widehat L_a=L_a/L_r,\quad
 \widehat\nu=\nu/(U_rL_r),\quad
 \widehat{\mathbf A}(\widehat t)=T_r\mathbf A(T_r\widehat t).
\end{gathered}
\label{eq:nondimensional_variables}
\end{equation}
Here $L_a$ are the computational periods. The same length scaling applies
to $\mathbf x'$, and $\widehat{\boldsymbol\kappa}=L_r\boldsymbol\kappa$.
The physical and transformed velocity coefficients are divided by $U_r$,
while $\widehat\phi=\phi/(\rev{\rho}U_r^2T_r)$. \rev{The mean pressure Hessian
and force gradient scale as $\widehat{\mathbf P}=T_r^2\mathbf P/\rho$ and
$\widehat{\mathbf G}=T_r^2\mathbf G$, respectively.}
The matrices $\mathbf B$, $\mathbf C_n$ and the factor $F_n$ are unchanged.

With these variables, Eq.~\eqref{eq:advancement_equations} becomes
\begin{equation}
\begin{aligned}
 \frac{d\widetilde{\widehat u}_i^{**}}{d\widehat t}
 &=-iF_n\widehat\tau_j
   \widetilde{\widehat u_i^*\widehat u_j^*}\\
 &\quad+i(C_n^{-1})_{ij}
   (\widehat A_{j\ell}-\widehat A_{\ell j})
   \widehat\kappa_\ell\widehat\phi,
\end{aligned}
\label{eq:nondimensional_advancement}
\end{equation}
where $\widehat{\boldsymbol\tau}=\mathbf C_n^{\mathsf T}
\widehat{\boldsymbol\kappa}$. The viscous factor is
\begin{equation}
 F_n=\exp\!\left[
 \frac{1}{Re_r}\int_{\widehat t_n}^{\widehat t}
 |\widehat{\boldsymbol\kappa}(s)|^2\,ds\right],
 \qquad Re_r=\frac{U_rL_r}{\nu}
 =\frac{K_0^2}{\nu\epsilon_0}.
\label{eq:nondimensional_viscous_factor}
\end{equation}
This choice gives $\widehat K_0=1$, $\widehat\epsilon_0=1$, and
$\widehat\nu=1/Re_r$. \rev{The coordinate and matrix update equations retain
their form. The pressure impulse scaling absorbs the density factors in the
dimensionless velocity reconstruction.}

\subsection{Periodic boundary conditions}
\label{sec:boundary_conditions}

We solve the transformed fluctuation equations on a fixed rectangular
computational domain in $\mathbf x'$, with period $L_j$ \rev{and an even
number $N_j$ of grid points} in direction
$\mathbf e_j$. With $\mathbf D=\operatorname{diag}(L_1,L_2,L_3)$, the
computational Fourier wavevectors are
$\mathbf k=2\pi\mathbf D^{-1}\mathbf n$, where
$\mathbf n\in\mathbb Z^3$ \rev{is the integer Fourier mode index. On the
discrete Fourier grid, its components satisfy $-N_j/2\le n_j<N_j/2$
for $j=1,2,3$}. Under $\mathbf x'=\mathbf B\mathbf x$, a
computational period $\mathbf D\mathbf e_j$ corresponds to the physical
translation

\begin{equation}
  \mathbf l_j(t)=\mathbf B^{-1}(t)\mathbf D\mathbf e_j.
  \label{eq:physical_periodic_basis}
\end{equation}

Thus the fluctuations remain periodic in the fixed computational domain,
while the physical lattice deforms.

\section{Numerical method}
\label{sec:numerical_method}

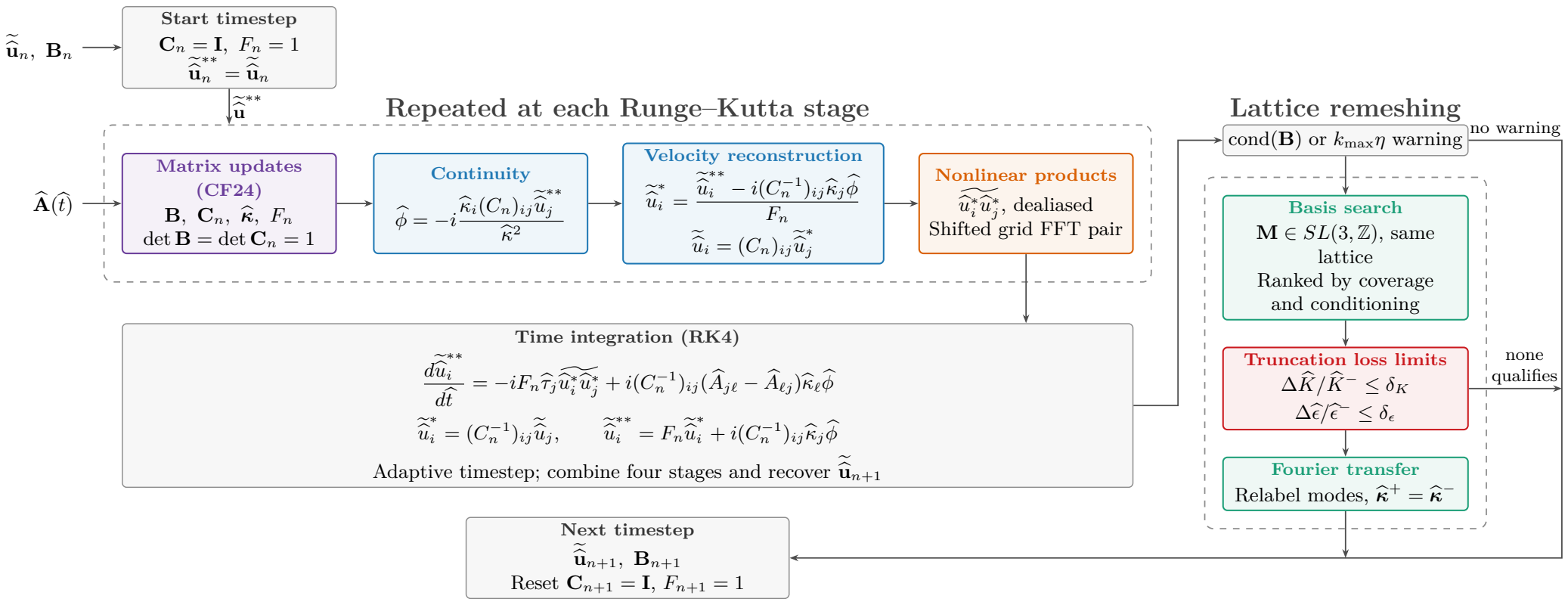
\begin{figure}[htbp]
\centering
\begingroup%
\hyphenpenalty=10000\relax%
\definecolor{dnsCoord}{HTML}{6A3D9A}%
\definecolor{dnsState}{HTML}{1F78B4}%
\definecolor{dnsNonlin}{HTML}{D95F02}%
\definecolor{dnsRemesh}{HTML}{1B9E77}%
\definecolor{dnsLoss}{HTML}{CB181D}%
\newcommand{\dnsfont}{\fontsize{9}{10.8}\selectfont}%
\newcommand{\dnstag}[2]{\textcolor{#1}{\fontsize{8}{9.6}\selectfont\bfseries #2}}%
\tikzset{%
 dnsmod/.style={draw=#1,fill=#1!7,line width=.7pt,rounded corners=2pt,%
   align=flush center,inner sep=2.5pt,font=\dnsfont},%
 dnsbox/.style={draw=black!45,fill=black!3,line width=.5pt,rounded corners=2pt,%
   align=flush center,inner sep=2.5pt,font=\dnsfont},%
 dnsarrow/.style={-{Stealth[length=4pt,width=3pt]},line width=.6pt,draw=black!70},%
 dnslabel/.style={font=\fontsize{8}{9.6}\selectfont,align=flush center,inner sep=1.5pt},%
 dnspart/.style={draw=black!45,dashed,line width=.6pt,rounded corners=4pt},%
 dnstitle/.style={font=\fontsize{12}{14.4}\selectfont\bfseries,text=black!75,%
   inner sep=1pt,align=flush center},%
}%
\sbox0{%
\begin{tikzpicture}[x=1cm,y=1cm,font=\dnsfont]
 \node[dnsbox,text width=3.25cm] (initialize) at (2.65,2.5)
   {\dnstag{black!80}{Start timestep}\\[1pt]
    $\mathbf C_n=\mathbf I,\ F_n=1$\\
    $\widetilde{\widehat{\mathbf u}}_n^{**}=\widetilde{\widehat{\mathbf u}}_n$};
 \node[anchor=east,align=center] (ain) at (0.3,2.5)
   {$\widetilde{\widehat{\mathbf u}}_n,\ \mathbf B_n$};
 \draw[dnsarrow] (ain) -- (initialize);
 \node[dnsmod=dnsCoord,text width=3.25cm,minimum height=1.6cm] (geometry) at (2.65,0)
   {\dnstag{dnsCoord}{Matrix updates (CF24)}\\[1pt]
    $\mathbf B,\ \mathbf C_n,\ \widehat{\boldsymbol\kappa},\ F_n$\\
    $\det\mathbf B=\det\mathbf C_n=1$};
 \node[dnsmod=dnsState,text width=3.25cm,minimum height=1.6cm] (continuity) at (6.68,0)
   {\dnstag{dnsState}{Continuity}\\[2pt]
    $\displaystyle\widehat\phi=-i\frac{\widehat\kappa_i(C_n)_{ij}
      \widetilde{\widehat u}_j^{**}}{\widehat\kappa^2}$};
 \node[dnsmod=dnsState,text width=4.0cm,minimum height=1.6cm] (reconstruct) at (11.05,0)
   {\dnstag{dnsState}{Velocity reconstruction}\\[2pt]
    $\displaystyle\widetilde{\widehat u}_i^*=
      \frac{\widetilde{\widehat u}_i^{**}-i(C_n^{-1})_{ij}\widehat\kappa_j\widehat\phi}{F_n}$\\[2pt]
    $\widetilde{\widehat u}_i=(C_n)_{ij}\widetilde{\widehat u}_j^*$};
 \node[dnsmod=dnsNonlin,text width=3.25cm,minimum height=1.6cm] (nonlinear) at (15.42,0)
   {\dnstag{dnsNonlin}{Nonlinear products}\\[1pt]
    $\widetilde{\widehat u_i^*\widehat u_j^*}$, dealiased\\
    Shifted grid FFT pair};
 \node[dnspart,fit=(geometry)(continuity)(reconstruct)(nonlinear),inner sep=8pt] (stages) {};
 \node[dnstitle,anchor=south] at (stages.north)
   {Repeated at each Runge--Kutta stage};
 \node[anchor=east] (meaninput) at (0.3,0) {$\widehat{\mathbf A}(\widehat t)$};
 \draw[dnsarrow] (meaninput) -- (geometry);
 \draw[dnsarrow] (initialize.south) -- (stages.north -| initialize.south)
   node[dnslabel,midway,right] {$\widetilde{\widehat{\mathbf u}}^{**}$};
 \draw[dnsarrow] (geometry) -- (continuity);
 \draw[dnsarrow] (continuity) -- (reconstruct);
 \draw[dnsarrow] (reconstruct) -- (nonlinear);

 \node[dnsbox,text width=16.02cm,anchor=north] (advance)
   at ($(stages.south)+(0,-.65)$)
   {\dnstag{black!80}{Time integration (RK4)}\\[2pt]
    $\dfrac{d\widetilde{\widehat u}_i^{**}}{d\widehat t}
     =-iF_n\widehat\tau_j\widetilde{\widehat u_i^*\widehat u_j^*}
     +i(C_n^{-1})_{ij}(\widehat A_{j\ell}-\widehat A_{\ell j})
       \widehat\kappa_\ell\widehat\phi$\\[4pt]
    $\widetilde{\widehat u}_i^*=(C_n^{-1})_{ij}\widetilde{\widehat u}_j,\qquad
     \widetilde{\widehat u}_i^{**}=F_n\widetilde{\widehat u}_i^*
       +i(C_n^{-1})_{ij}\widehat\kappa_j\widehat\phi$\\[4pt]
    Adaptive timestep; combine four stages and recover $\widetilde{\widehat{\mathbf u}}_{n+1}$};
 \draw[dnsarrow] (nonlinear) -- (advance.north -| nonlinear);

 \node[dnsbox,text width=3.75cm,anchor=north] (trigger)
   at (20.54,0 |- stages.north)
   {$\operatorname{cond}(\mathbf B)$ or $k_{\max}\eta$ warning};
 \node[dnsmod=dnsRemesh,text width=3.75cm,anchor=north] (search)
   at ($(trigger.south)+(0,-.62)$)
   {\dnstag{dnsRemesh}{Basis search}\\[1pt]
    $\mathbf M\in SL(3,\mathbb Z)$, same lattice\\
    Ranked by coverage\\ and conditioning};
 \node[dnsmod=dnsLoss,text width=3.75cm,anchor=north] (loss)
   at ($(search.south)+(0,-.42)$)
   {\dnstag{dnsLoss}{Truncation loss limits}\\[1pt]
    $\Delta\widehat K/\widehat K^-\le\delta_K$\\[2pt]
    $\Delta\widehat\epsilon/\widehat\epsilon^-\le\delta_\epsilon$};
 \node[dnsmod=dnsRemesh,text width=3.75cm,anchor=north] (transfer)
   at ($(loss.south)+(0,-.42)$)
   {\dnstag{dnsRemesh}{Fourier transfer}\\[1pt]
    Relabel modes, $\widehat{\boldsymbol\kappa}^+=\widehat{\boldsymbol\kappa}^-$};
 \node[dnspart,fit=(search)(loss)(transfer),inner sep=8pt] (remesh) {};
 \node[dnstitle,anchor=south] at (trigger.north) {Lattice remeshing};
 \draw[dnsarrow] (trigger) -- (search);
 \draw[dnsarrow] (search) -- (loss);
 \draw[dnsarrow] (loss) -- (transfer);
 \coordinate (corridor) at ($(stages.east)!.5!(remesh.west)$);
 \draw[dnsarrow] (advance.east) -- (corridor |- advance.east) |- (trigger.west);
 \coordinate (outputlevel) at ($(transfer.south)+(0,-.75)$);
 \node[dnsbox,text width=5.0cm] (next) at (9.035,0 |- outputlevel)
   {\dnstag{black!80}{Next timestep}\\[1pt]
    $\widetilde{\widehat{\mathbf u}}_{n+1},\ \mathbf B_{n+1}$\\
    Reset $\mathbf C_{n+1}=\mathbf I$, $F_{n+1}=1$};
 \coordinate (bypass) at ($(remesh.east)+(1.2,0)$);
 \draw[dnsarrow] (trigger.east) -- (trigger.east -| bypass)
   node[dnslabel,midway,above] {no warning}
   |- (next.east);
 \draw[dnsarrow] (loss.east) -- (loss.east -| bypass)
   node[dnslabel,pos=.6,above] {none\\ qualifies};
 \draw[dnsarrow] (transfer.south) -- (transfer.south |- outputlevel);
\end{tikzpicture}}%
\ifdim\wd0>\linewidth\relax\resizebox{\linewidth}{!}{\usebox0}\else\usebox0\fi%
\endgroup%
\caption{Solver algorithm. CF24 matrix updates, continuity enforcement, velocity reconstruction, and nonlinear products supply each RK4 stage
derivative. The combined endpoint is reconstructed and checked before
optional lattice remeshing. The resulting physical state starts the next
timestep.}
\label{fig:solver_algorithm}
\end{figure}

Figure~\ref{fig:solver_algorithm} summarizes the algorithm. Each timestep
starts from the physical field and coordinate map at $t_n$. We first reset the
step-local transformations and initialize
$\widetilde{\mathbf u}_n^{**}=\widetilde{\mathbf u}_n$.
At each Runge--Kutta stage, the fourth-order, two-exponential
commutator-free Magnus integrator (CF24) evaluates the coordinate and
velocity maps from this starting state while preserving their unit determinants.
The resulting coordinate map gives the physical wavevectors
$\boldsymbol\kappa$, which we use to evaluate the viscous factor $F_n$
from the time integral of $\kappa^2$ [Eq.~\eqref{eq:advancement_variables}].
Continuity then determines the pressure impulse from the double-star
coefficients, allowing us to reconstruct the single-star velocity for the
nonlinear products and the physical velocity for the stability checks.
We evaluate these products using Patterson--Orszag truncation to remove
double and triple aliases, and \rev{fast Fourier transforms (FFTs) on paired
shifted grids} to cancel the remaining single aliases. Once all stage
derivatives are available, the \rev{classical fourth-order Runge--Kutta
method (RK4)} combines them and we reconstruct the endpoint velocity.
Together, the stage and endpoint stability checks control the timestep.
When remeshing is required, we search over integer changes of lattice basis
and apply the spectral coverage, conditioning and truncation loss criteria.
We then reindex the retained Fourier modes in the selected basis.
The resulting field and coordinate map initialize the next timestep.

\subsection{Unit determinant preserving matrix updates}

\rev{Since Runge--Kutta updates can introduce determinant drift
\citep{girimajiPope1990},
we use the two-exponential CF24 scheme of \citet{blanesMoan2006} to preserve
unit determinants while retaining fourth order global accuracy.}
For a timestep $h$ beginning at $t_n$, define

\begin{equation}
\begin{aligned}
  c_\pm&=\frac12\pm\frac{\sqrt3}{6},
  &a_\pm&=\frac14\pm\frac{\sqrt3}{6},\\
  \mathbf A_\pm&=\mathbf A(t_n+c_\pm h),
  &\mathbf D_1&=a_+\mathbf A_-+a_-\mathbf A_+,\\
  \mathbf D_2&=a_-\mathbf A_-+a_+\mathbf A_+.
\end{aligned}
  \label{eq:cf24_generators}
\end{equation}

The updates are

\begin{equation}
  \mathbf B(t_n+h)\approx\mathbf B(t_n)
    e^{-h\mathbf D_1}e^{-h\mathbf D_2},
  \qquad
  \mathbf C_n(t_n+h)\approx
    e^{-h\mathbf D_2}e^{-h\mathbf D_1}.
  \label{eq:cf24_updates}
\end{equation}

The factor order is reversed because the matrix equations in
Eq.~\eqref{eq:map_evolution} act on opposite sides. This preserves the
ordering of successive, potentially noncommuting mean velocity gradients.
Since $\operatorname{tr}\mathbf A(t)=0$, each $\mathbf D_r$ is trace
free and its exponential has determinant one. Both maps therefore
preserve unit determinant by construction.

\subsection{Time integration of the fluctuations}

Using these maps, we advance the double-star equation,
Eq.~\eqref{eq:advancement_equations}, with the classical four-stage,
fourth-order Runge--Kutta method. At each
stage time, $\mathbf B$, $\mathbf C_n$, $\boldsymbol\kappa$, and $F_n$ are
evaluated over the elapsed interval from the accepted state at $t_n$.
The two midpoint stages share the same geometry but have different
double-star fields.
The viscous integral in Eq.~\eqref{eq:advancement_variables} is evaluated
by two-point Gauss quadrature. For an elapsed time $s$,

\begin{equation}
  \log F_n(\mathbf k,t_n+s)
  \approx\frac{\nu s}{2}\left[
    \kappa^2(\mathbf k,t_n+c_-s)
    +\kappa^2(\mathbf k,t_n+c_+s)\right].
  \label{eq:viscous_gauss_quadrature}
\end{equation}

The coordinate maps at the quadrature points are computed with CF24.
At each Runge--Kutta stage, continuity gives $\phi$, and the recovered
single-star velocity is used to evaluate the nonlinear products. After the four
stage derivatives are combined, $\phi$ is recomputed from the final
double-star field and the physical velocity is recovered.

The timestep is constrained by fluctuation advection on the fixed
computational grid. With $N_i$ points and spacing
$\Delta x'_i=L_i/N_i$ in direction $i$, the
\rev{Courant--Friedrichs--Lewy (CFL) condition at each stage} is

\begin{equation}
  h\max_{\mathbf x'}\sum_i
  \frac{|(\mathbf B\mathbf u)_i|}{\Delta x'_i}
  \le C_{\rm CFL}.
  \label{eq:computational_cfl}
\end{equation}

We check this condition on each reconstructed stage field and the
final field. Additional bounds on $h\|\mathbf A\|_2$ and
$h^2\|\dot{\mathbf A}\|_2$ restrict the timestep according to the magnitude
and rate of change of the mean velocity gradient. These bounds are estimated from samples over
each trial interval. A failed trial is repeated with a shorter timestep
from the preceding accepted state.
The simulations use $C_{\rm CFL}=0.4$, with bounds
$h\|\mathbf A\|_2\le2.3\times10^{-3}$ and
$h^2\|\dot{\mathbf A}\|_2\le5\times10^{-3}$.
We selected these values based on preliminary simulations.

\subsection{Spectral discretization and dealiasing}

The transformed fluctuations are represented on a fixed
$N_1\times N_2\times N_3$ computational grid. We retain the integer
mode indices

\begin{equation}
  \mathcal K_N=\left\{\mathbf n\in\mathbb Z^3:
  \sum_{a=1}^3\left(\frac{n_a}{N_a}\right)^2<\frac{2}{9},
  \quad |n_a|<\frac{N_a}{2}\ \text{for }a=1,2,3\right\}.
  \label{eq:retained_spectral_sphere}
\end{equation}

For $N_1=N_2=N_3=N$, Eq.~\eqref{eq:retained_spectral_sphere} reduces to
the Patterson--Orszag sphere
$\|\mathbf n\|_2<\sqrt{2}N/3$
\citep{pattersonOrszag1971,rogallo1981}.
This truncation removes double and triple aliases.
\rev{At each Runge--Kutta stage, we use phase shifts and inverse FFTs to
reconstruct the same single-star velocity field on two computational grids
offset by half the grid spacing in each direction. On each grid, we form the
products $u_i^*u_j^*$ pointwise, transform them back to Fourier space and
undo the corresponding phase shift. Averaging the two sets of coefficients
cancels the remaining single aliases \citep{rogallo1981}.}

\subsection{General three-dimensional remeshing}
\label{sec:remeshing}

Between remeshes, each Fourier coefficient retains its mode index
$\mathbf n$, while its physical wavevector
$\boldsymbol\kappa=\mathbf B^{\mathsf T}\mathbf k$ changes with the mean
deformation. Consequently, the finite set of represented physical
wavevectors can become distorted and lose resolution in some
directions. Because the advancement equation,
Eq.~\eqref{eq:advancement_equations}, depends on $\boldsymbol\kappa$,
this loss of resolution can affect the computed fluctuations.

To address these issues, remeshing selects an equivalent basis of the same
deformed physical lattice using physical spectral coverage and the condition
number of $\mathbf B$. \rev{The condition number
$\operatorname{cond}_2(\mathbf B)=\sigma_{\max}(\mathbf B)/\sigma_{\min}(\mathbf B)$
is the ratio of the largest to the smallest singular value of $\mathbf B$ and
measures how unevenly the coordinate map stretches different directions.}
The coefficients are then reindexed within the existing computational mode set.
During this transfer, each retained mode keeps its physical wavevector
and velocity coefficient, while modes mapped outside the retained set
are discarded and unfilled indices are set to zero.
For simple shear, the basis change can be prescribed analytically
\citep{rogallo1981}. In the general case, we search over integer
changes of basis.

\subsubsection{Periodic lattice and integer changes of basis}

Following \citet{dobson2014}, we index periodic replicas by integer triples
$\mathbf m\in\mathbb Z^3$. In deforming coordinates, a point $\mathbf x'$
has replicas at $\mathbf x'+\mathbf D\mathbf m$, where $\mathbf D$ is the
diagonal matrix of computational domain lengths defined in
Section~\ref{sec:boundary_conditions} and $m_a$ counts the signed number of
periods along direction $a$. The map
$\mathbf x=\mathbf B^{-1}\mathbf x'$ then gives the physical translation
$\mathbf B^{-1}\mathbf D\mathbf m$. Thus the physical lattice basis and its
set of periodic translations are

\begin{equation}
  \mathbf H=\mathbf B^{-1}\mathbf D,
  \qquad
  \Lambda(\mathbf H)
  =\{\mathbf H\mathbf m:\mathbf m\in\mathbb Z^3\}.
  \label{eq:remesh_lattice}
\end{equation}

The columns $\mathbf h_j$ of $\mathbf H$ are the vectors in
Eq.~\eqref{eq:physical_periodic_basis}. To remap the lattice, let $\mathbf M$
denote the change of basis. Superscripts $-$ and $+$ denote states immediately
before and after a remesh. Keeping $\mathbf D$ unchanged, we substitute
$\mathbf H=\mathbf B^{-1}\mathbf D$ into the basis change and invert to obtain
the coordinate map.

\begin{equation}
\begin{aligned}
  \mathbf H^+&=\mathbf H^-\mathbf M,\\
  (\mathbf B^+)^{-1}\mathbf D
  &=(\mathbf B^-)^{-1}\mathbf D\mathbf M,\\
  \mathbf B^+&=\mathbf D\mathbf M^{-1}\mathbf D^{-1}\mathbf B^-.
\end{aligned}
  \label{eq:remesh_basis_change}
\end{equation}

The physical periodic lattice is preserved if
$\Lambda(\mathbf H^+)=\Lambda(\mathbf H^-)$. Using
Eq.~\eqref{eq:remesh_basis_change}, we write the new lattice as
\begin{equation}
\begin{aligned}
  \Lambda(\mathbf H^+)
  &=\{\mathbf H^+\mathbf m:\mathbf m\in\mathbb Z^3\}\\
  &=\{\mathbf H^-\mathbf M\mathbf m:\mathbf m\in\mathbb Z^3\}.
\end{aligned}
  \label{eq:remesh_new_lattice}
\end{equation}
To ensure that every new lattice point belongs to the original lattice,
$\mathbf M\mathbf m$ must be an integer vector for every
$\mathbf m\in\mathbb Z^3$, so $\mathbf M$ must have integer entries.
The two bases must also give boxes of equal volume. Since the basis change
scales the box volume by $|\det\mathbf M|$, we require $|\det\mathbf M|=1$.

Conversely, suppose $\mathbf M$ has integer entries and $|\det\mathbf M|=1$.
The adjugate has integer entries, so
$\mathbf M^{-1}=\operatorname{adj}(\mathbf M)/\det\mathbf M$ does as well.
It follows that for each old lattice label $\mathbf m^-\in\mathbb Z^3$,
the new label $\mathbf m^+=\mathbf M^{-1}\mathbf m^-$ is an integer
vector and satisfies
\begin{equation}
  \mathbf H^-\mathbf m^-=\mathbf H^+\mathbf m^+
  =\mathbf H^-\mathbf M\mathbf m^+.
  \label{eq:remesh_lattice_label}
\end{equation}
Since $\mathbf M$ is invertible, the new label is unique. The integer
label map is therefore bijective, so $\mathbf M\mathbb Z^3=\mathbb Z^3$
and $\Lambda(\mathbf H^+)=\Lambda(\mathbf H^-)$.

We also require the coordinate map to preserve its unit determinant.
Since $\det\mathbf B^+=\det\mathbf B^-/\det\mathbf M$, this requires
$\det\mathbf M=1$. Thus the changes of basis that preserve both the
lattice and the unit determinant of $\mathbf B$ are exactly those
in $SL(3,\mathbb Z)$.

The computational coordinates and mode indices change consistently according to
$\mathbf x'^+=\mathbf D\mathbf M^{-1}\mathbf D^{-1}\mathbf x'^-$ and
$\mathbf n^+=\mathbf M^{\mathsf T}\mathbf n^-$.
Hence $\mathbf k^+=\mathbf D^{-1}\mathbf M^{\mathsf T}
\mathbf D\mathbf k^-$, and

\begin{equation}
  \boldsymbol\kappa^+
  =(\mathbf B^+)^{\mathsf T}\mathbf k^+
  =(\mathbf B^-)^{\mathsf T}\mathbf k^-
  =\boldsymbol\kappa^-.
  \label{eq:remesh_wavevector_invariance}
\end{equation}

The Fourier phase and physical velocity field are preserved for modes
that survive the spectral truncation. The lattice transformation and Fourier
relabeling are also derived in \citet{hajiyev2026rogallo}.

\subsubsection{Selection of the new lattice basis}

To compare candidate bases, we measure physical spectral coverage by the
smallest radius of the retained ellipsoid in physical wavevector space,

\begin{equation}
 k_{\max}=\frac{\sqrt{2}}{3}\,
 \sigma_{\min}\!\left(\mathbf B^{\mathsf T}
 \operatorname{diag}\!\left(
 \frac{2\pi N_1}{L_1},\frac{2\pi N_2}{L_2},\frac{2\pi N_3}{L_3}
 \right)\right),
 \qquad \eta=(\nu^3/\epsilon)^{1/4},
 \label{eq:physical_resolution}
\end{equation}
where $\sigma_{\min}(\cdot)$ denotes the smallest singular value of the
complete matrix product in parentheses.
We use $k_{\max}\eta$ to assess resolution in the least covered physical
direction. We also report $k_{\max,\mathrm{comp}}\eta$ for the computational
grid, where
$k_{\max,\mathrm{comp}}=(\sqrt{2}/3)\min_a(2\pi N_a/L_a)$.
Both estimates use the physical Kolmogorov length $\eta$.
A search is triggered when $\operatorname{cond}_2(\mathbf B)>1.5$ or
$k_{\max}\eta<1.5$, with repeated searches every $20$ accepted steps
under persistent warnings.

To represent the same physical lattice in a new basis, we use special linear
matrices with integer entries and determinant one. Elementary integer column
operations preserve these properties, so their compositions generate candidate
bases from $\mathbf H$. Each operation replaces $\mathbf h_j$ by
$\mathbf h_j-a\mathbf h_i$. For each pair, the nearest-integer coefficient
minimizes the new column length, while $a=\pm1$ also explores simple column
additions and subtractions. Algorithm~\ref{alg:remeshing_search} ranks these bases
by decreasing physical cutoff, preferring smaller condition numbers when
cutoffs are equal. Final selection requires the joint improvement rule:
neither measure worsens and at least one improves beyond roundoff.
Predicted truncation losses must also satisfy the limits in the following
subsection.
In Algorithm~\ref{alg:remeshing_search}, $k_{\max}(\mathbf B)$ is evaluated
from Eq.~\eqref{eq:physical_resolution}, and $\operatorname{nint}$ denotes
rounding to the nearest integer. The reported simulations use search depth
$d=3$ and extend up to $w=8$ candidates per level.

\begin{algorithm}
\hrule
\smallskip
\footnotesize
\noindent\textbf{Input:} Coordinate map $\mathbf B$, physical Fourier coefficients
$\widetilde{\mathbf u}$, grid $N_a,L_a$, search parameters $d,w$ and loss limits
$\delta_K,\delta_\epsilon$.\\
\textbf{Output:} Selected basis change $\mathbf M$, defining $\mathbf H^+=\mathbf H\mathbf M$.
\begin{tabbing}
\hspace{1em}\=\hspace{1em}\=\hspace{1em}\=\hspace{1em}\=\kill
$\mathbf H\gets\mathbf B^{-1}\mathbf D$;
$k_0\gets k_{\max}(\mathbf B)$;
$c_0\gets\operatorname{cond}_2(\mathbf B)$ \\
$\mathcal B\gets\{\mathbf I\}$;
$\mathcal P\gets\varnothing$ \quad (current search matrices and all candidates) \\
\textbf{for} $\ell=1,\ldots,d$ \textbf{do} \\
\>$\mathcal Q\gets\varnothing$ \\
\>\textbf{for each} $\mathbf M\in\mathcal B$ and ordered pair $i\ne j$ \textbf{do} \\
\>\>$[\mathbf h_1,\mathbf h_2,\mathbf h_3]\gets\mathbf H\mathbf M$ \\
\>\>$a_*\gets(\mathbf h_i\cdot\mathbf h_j)/|\mathbf h_i|^2$ \\
\>\>\textbf{for each} $a\in\{-1,1,\operatorname{nint}(a_*)\}\setminus\{0\}$ \textbf{do} \\
\>\>\>$\mathcal Q\gets\mathcal Q\cup\{\mathbf M(\mathbf I-a\mathbf e_i\mathbf e_j^{\mathsf T})\}$ \\
\>\>\textbf{end for} \\
\>\textbf{end for} \\
\>$\mathcal Q\gets\mathcal Q\setminus(\mathcal P\cup\{\mathbf I\})$ \\
\>\textbf{for each} $\mathbf M\in\mathcal Q$ \textbf{do} \\
\>\>$\mathbf B_{\mathbf M}\gets\mathbf D\mathbf M^{-1}\mathbf D^{-1}\mathbf B$ \\
\>\>$k_{\mathbf M}\gets k_{\max}(\mathbf B_{\mathbf M})$;
$c_{\mathbf M}\gets\operatorname{cond}_2(\mathbf B_{\mathbf M})$ \\
\>\textbf{end for} \\
\>$\mathcal P\gets\mathcal P\cup\mathcal Q$ \\
\>Rank $\mathcal Q$ by decreasing $k_{\mathbf M}$, then increasing $c_{\mathbf M}$ \\
\>$\mathcal B\gets$ up to $w$ best matrices from $\mathcal Q$ \\
\textbf{end for} \\
Filter $\mathcal P$ using the joint improvement rule relative to $(k_0,c_0)$ \\
Rank $\mathcal P$ by decreasing $k_{\mathbf M}$, then increasing $c_{\mathbf M}$ \\
\textbf{for each} $\mathbf M\in\mathcal P$ \textbf{do} \\
\>Predict $\Delta K/K^-$ and $\Delta\epsilon/\epsilon^-$ from
Eq.~\eqref{eq:remesh_truncation_losses} \\
\>\textbf{if} $\Delta K/K^-\le\delta_K$ \textbf{and}
$\Delta\epsilon/\epsilon^-\le\delta_\epsilon$ \textbf{then} \\
\>\>\textbf{return} $\mathbf M$ \\
\>\textbf{end if} \\
\textbf{end for} \\
\textbf{return} $\mathbf I$
\end{tabbing}
\smallskip
\hrule
\caption{Candidate lattice basis search and selection.}
\label{alg:remeshing_search}
\end{algorithm}

\subsubsection{Fourier transfer and truncation losses}

During remeshing, the relabeling
$\mathbf n^+=\mathbf M^{\mathsf T}\mathbf n^-$ is applied to each retained
mode. When the new index remains in the spectral set of
Eq.~\eqref{eq:retained_spectral_sphere}, its physical Cartesian velocity
coefficient is copied:

\begin{equation}
  \widetilde u_i^+(\mathbf k^+)
  =\widetilde u_i^-(\mathbf k^-).
  \label{eq:remesh_coefficient_transfer}
\end{equation}

The velocity components are not rotated or interpolated. Modes whose
new indices fall outside the spectral set are discarded rather than
wrapped into it, and unfilled coefficients are set to zero.
To quantify these truncation losses, let $\mathcal L$ be the old modes
that a candidate basis would discard. The resulting energy and dissipation
losses are

\begin{equation}
  \Delta K=\frac12\sum_{\mathbf n\in\mathcal L}\sum_i|\widetilde u_i^-|^2,
  \qquad
  \Delta\epsilon=\nu\sum_{\mathbf n\in\mathcal L}
    |\boldsymbol\kappa^-|^2\sum_i|\widetilde u_i^-|^2.
  \label{eq:remesh_truncation_losses}
\end{equation}

A small energy loss alone does not limit the removal of dissipative
small-scale modes. We therefore require

\begin{equation}
  \frac{\Delta K}{K^-}\le\delta_K,
  \qquad
  \frac{\Delta\epsilon}{\epsilon^-}\le\delta_\epsilon,
  \label{eq:remesh_loss_acceptance}
\end{equation}
where $K^-$ and $\epsilon^-$ are the totals before remeshing and
$\delta_K=10^{-3}$ and $\delta_\epsilon=10^{-2}$ are the loss limits.

\section{Initial isotropic fields}
\label{sec:initial_fields}

We develop the initial isotropic fields from divergence-free random
velocity fields with square-pulse energy spectra in the forcing band.
After each timestep, we rescale the Fourier coefficients in that band by a
common factor to restore the target kinetic energy. Fields used for mean
deformation are subsequently advanced without this forcing.
All five fields use a cubic domain of side $2\pi$ and $\nu=0.004299$.
Forcing bands refer to $|\boldsymbol\kappa|/(2\pi/L_{\min})$, where
$L_{\min}=\min_j L_j$.

We calculate skewness using the generalized definition of \citet{mansourCambonSpeziale1991}, with the opposite sign to
follow the negative convention of \citet{blaisdellShariff1996}:
\begin{equation}
 S=-\frac{6\sqrt{15}}{7}
 \frac{\left\langle\omega_i
 \dfrac{\partial u_i}{\partial x_j}\omega_j\right\rangle}
 {\langle\omega_i\omega_i\rangle^{3/2}}.
 \label{eq:generalized_derivative_skewness}
\end{equation}
Here $\boldsymbol\omega=\nabla\times\boldsymbol u$ is the fluctuating
vorticity and derivatives are taken in physical coordinates. Under
isotropy, this diagnostic reduces to the longitudinal velocity derivative
skewness,
\begin{equation}
 S=-S_{\mathrm{MW}}
 =\frac{\langle u_{1,1}^{3}\rangle}
 {\langle u_{1,1}^{2}\rangle^{3/2}},
 \qquad u_{1,1}=\frac{\partial u_1}{\partial x_1},
 \label{eq:skewness_isotropic_limit}
\end{equation}
where $S_{\mathrm{MW}}$ is the negative derivative skewness defined by
\citet{mansour1994}.
\rev{In developed isotropic turbulence, the skewness of the longitudinal
velocity derivative is typically close to $-0.5$ \citep{blaisdellShariff1996}.}

Following \citet{pope2000}, the scalar velocity scale is
\begin{equation}
 u'=\sqrt{\frac{2K}{3}},
 \label{eq:rms_velocity_definition}
\end{equation}
with Taylor microscale
\begin{equation}
 \lambda=\sqrt{\frac{15\nu u'^2}{\epsilon}}
 \label{eq:taylor_microscale_definition}
\end{equation}
and Reynolds number
\begin{equation}
 Re_\lambda=\frac{u'\lambda}{\nu}.
 \label{eq:taylor_reynolds_definition}
\end{equation}
Here $\lambda$ corresponds to Pope's $\lambda_g$. These isotropic definitions
also provide scalar diagnostics after anisotropy develops.
Using the isotropic spectral relations, we estimate the longitudinal
integral scale of the initial fields by the discrete Fourier sum
\begin{equation}
 L_{\mathrm{int}}=\frac{3\pi}{8K}
 \sum_{\boldsymbol\kappa\ne\boldsymbol 0}
 \frac{\sum_i|\widetilde u_i(\boldsymbol\kappa)|^2}
 {|\boldsymbol\kappa|},
 \label{eq:integral_scale_definition}
\end{equation}
where the Fourier coefficients are normalized so that
$K=\frac12\sum_{\boldsymbol\kappa}\sum_i|\widetilde u_i|^2$.
This estimate corresponds to the longitudinal integral scale under
isotropy; after anisotropy develops, it serves as a scalar spectral
length scale diagnostic. It is invariant under exact remeshing.

The Reynolds stress anisotropy tensor is
\begin{equation}
 b_{ij}=\frac{\langle u_i u_j\rangle}{2K}-\frac{\delta_{ij}}{3},
 \label{eq:anisotropy_tensor_definition}
\end{equation}
and its invariants are
\begin{equation}
 (I_b,II_b,III_b)=\left(b_{ii},-\frac12 b_{ij}b_{ji},
 \frac13 b_{ij}b_{jk}b_{ki}\right),
 \label{eq:anisotropy_invariant_definition}
\end{equation}
following \citet{pope2000}.

To assess the development of the initial isotropic fields, we use the $3$--$6$
forcing band on a $336^3$ grid at $Re_\lambda=89.530$ and on $256^3$ grids at
$Re_\lambda=84.153$ and $47.507$ (Table~\ref{tab:initial_fields}).
To select the $336^3$ field, we compare three block means
over a trailing window of $0.5$ simulation units. The relative spread,
defined as the range of block means divided by their mean, must be below
$2\%$ for $Re_\lambda$, dissipation, and integral scale in two successive
checks, performed every $100$ accepted timesteps.
When this criterion is met, we stop the forced simulation and accept the
final field as the initial isotropic field.

For the general mean deformation cases, we select $256^3$ and $128^3$
fields at $Re_\lambda=33.211$ and $15.207$, respectively, using the
$8$--$16$ forcing band.
Both fields pass two consecutive checks using the same block-mean screening
procedure and $2\%$ spread threshold. Their trailing window covers two
turnover times, $\int\epsilon/(2K)\,dt=2$.
The largest block mean spreads are $0.94\%$ and $0.45\%$, respectively.

\begin{table}[!ht]
\centering
\fontsize{9.5}{11.5}\selectfont
\setlength{\tabcolsep}{2.5pt}
\renewcommand{\arraystretch}{1.10}
\begin{tabular*}{\linewidth}{@{\extracolsep{\fill}}lccccc@{}}
\toprule
Quantity & $Re_\lambda=89.530$ & $Re_\lambda=84.153$ & $Re_\lambda=47.507$ & $Re_\lambda=33.211$ & $Re_\lambda=15.207$ \\
\midrule
Grid & $336^3$ & $256^3$ & $256^3$ & $256^3$ & $128^3$ \\
Forcing band & $3$--$6$ & $3$--$6$ & $3$--$6$ & $8$--$16$ & $8$--$16$ \\
\addlinespace[3pt]
$K$ & $11.5500$ & $8.9713$ & $1.1711$ & $2.5000$ & $0.2500$ \\
$\epsilon$ & $25.8089$ & $17.6242$ & $0.9424$ & $8.7874$ & $0.4191$ \\
$S$ & $-0.5034$ & $-0.4853$ & $-0.5063$ & $-0.4893$ & $-0.4652$ \\
\addlinespace[3pt]
\rev{$\langle u_1u_1\rangle/(2K)$} & $0.3363$ & $0.3259$ & $0.3317$ & $0.3378$ & $0.3280$ \\
\rev{$\langle u_2u_2\rangle/(2K)$} & $0.3172$ & $0.3226$ & $0.3336$ & $0.3323$ & $0.3338$ \\
\rev{$\langle u_3u_3\rangle/(2K)$} & $0.3465$ & $0.3515$ & $0.3346$ & $0.3299$ & $0.3382$ \\
\addlinespace[3pt]
$I_b$ & $4.44\times10^{-16}$ & $-1.05\times10^{-15}$ & $-1.42\times10^{-14}$ & $-4.50\times10^{-15}$ & $-1.11\times10^{-16}$ \\
$II_b$ & $-5.94\times10^{-4}$ & $-3.96\times10^{-4}$ & $-1.77\times10^{-4}$ & $-2.13\times10^{-5}$ & $-3.04\times10^{-5}$ \\
$III_b$ & $4.02\times10^{-6}$ & $2.47\times10^{-6}$ & $-4.17\times10^{-7}$ & $9.51\times10^{-9}$ & $-2.54\times10^{-8}$ \\
\addlinespace[3pt]
$k_{\max}\eta$ & $1.180$ & $0.989$ & $2.056$ & $1.177$ & $1.259$ \\
$L_{\mathrm{int}}/L_{\min}$ & $0.0638$ & $0.0647$ & $0.0716$ & $0.0293$ & $0.0356$ \\
\bottomrule
\end{tabular*}
\caption{Selected initial isotropic fields, ordered by decreasing $Re_\lambda$.
The first three fields assess the forcing procedure, and the final two
initialize the general mean deformation cases.
All use $\nu=0.004299$ and a cubic domain of side $2\pi$.
Here $k_{\max}\eta$ uses the physical cutoff.}
\label{tab:initial_fields}
\end{table}

\begin{figure}[htbp]
\centering
\includegraphics[width=\linewidth]{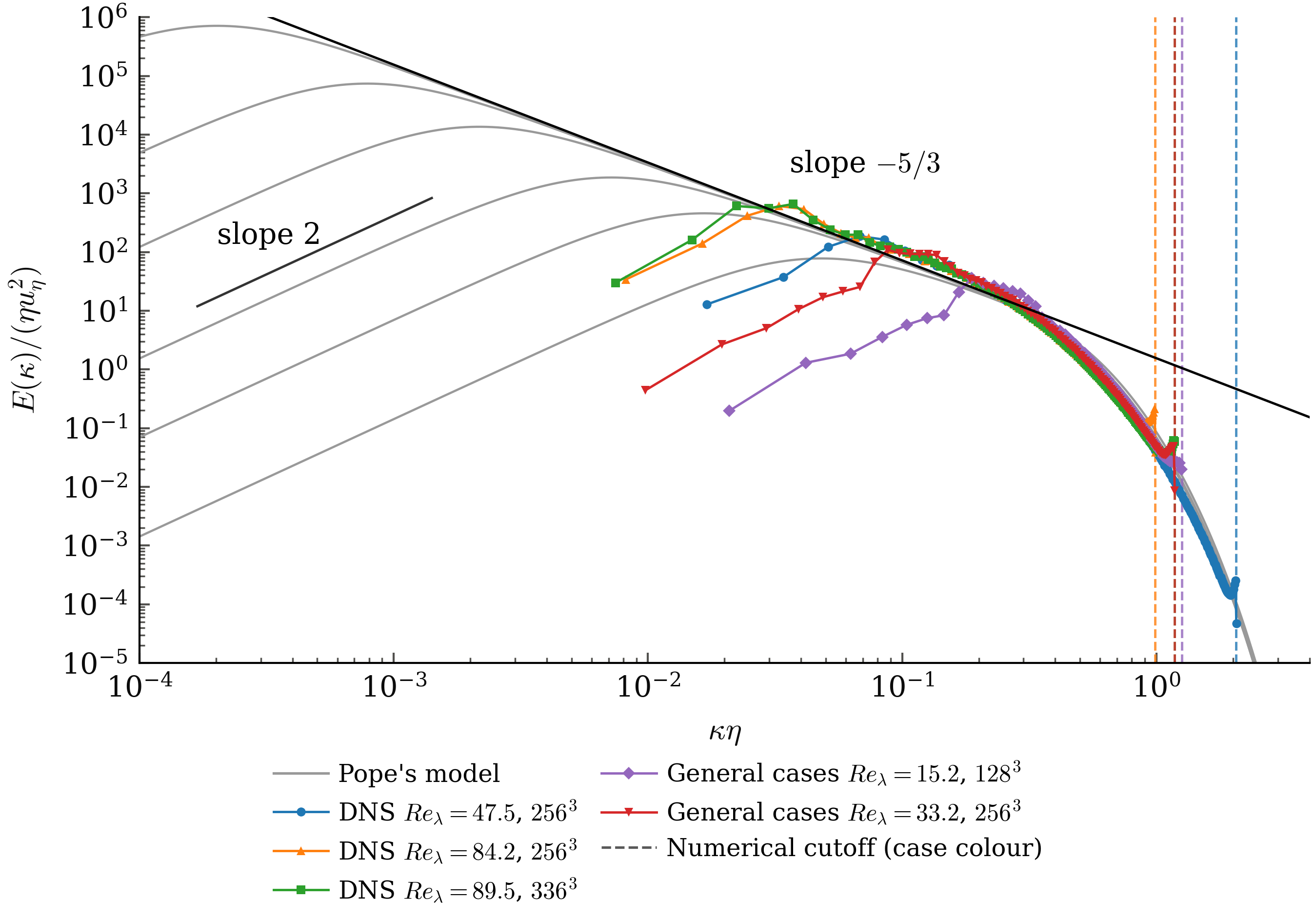}
\caption{Three-dimensional energy spectra in Kolmogorov units. Coloured
symbols show spectra from the three fields used to assess the forcing procedure
and the two initial isotropic fields for the general deformation cases.
Grey curves show evaluations of
\rev{the model spectrum given by \citet{pope2000}} at
$Re_\lambda=30,70,130,300,600,$ and $1500$. Dashed vertical lines mark
the numerical cutoff for each case in the corresponding colour. Solid
guides indicate slopes of $2$ and $-5/3$.}
\label{fig:initial_field_spectra}
\end{figure}

Figure~\ref{fig:initial_field_spectra} compares the spectra of these fields
with \rev{the model spectrum given by \citet{pope2000}} in Kolmogorov
units. The spectral scale on
the vertical axis is $\eta u_\eta^2=(\epsilon\nu^5)^{1/4}$, where
$u_\eta=(\nu\epsilon)^{1/4}$ and $\eta$ is defined in
Eq.~\eqref{eq:physical_resolution}. The model curves are evaluated at the stated
Reynolds numbers, with their coefficients determined from the energy and
dissipation integrals. The near-cutoff upturn is consistent with the
truncation effects reported by \citet{jangDeBruynKops2007}.

\section{Validation}
\label{sec:validation}

For the validation cases, we follow the reference studies' initialization
procedures and prescribed mean velocity gradients as closely as possible
\citep{leeReynolds1985,blaisdellShariff1996}.

\subsection{Lee--Reynolds benchmarks}

We compare three isotropic decay cases and fourteen primary deformations
from \citet{leeReynolds1985} (Table~\ref{tab:lee_reynolds_main}).
The latter comprise eight plane strains,
three axisymmetric contractions, and three axisymmetric expansions,
all with constant mean velocity gradients and no mean rotation. The simulations
use the reference $128^3$ grid, case-specific rectangular box dimensions,
and imposed strain rates. Each family shares one independently generated
initial field: the plane strain, contraction, and expansion cases use
fields developed from HIA, HIC, and HID, respectively. Lee--Reynolds use
\rev{a second-order Runge--Kutta method}, whereas the present solver uses
RK4 with CF24 matrix updates.
Each primary deformation undergoes one remesh, consisting of a signed
permutation of the coordinate axes. During these remeshes, energy and
dissipation are preserved to roundoff.

\begin{table}[!htbp]
\centering
\footnotesize
\setlength{\tabcolsep}{3pt}
\begin{tabular}{lllccccc}
\toprule
& \multicolumn{2}{c}{\rev{$Re_\lambda$ range}} & $c_b$ & $10^2(-II_b)$ & $c_E^{\dagger}$ & $(K/K_0)^{\dagger}$ & Max. error (\%)$^{\dagger}$ \\
\cmidrule(lr){2-3}
Case & Reference & Present & & Ref./present & & Ref./present & Energy \\
\midrule
\addlinespace[2pt]
\multicolumn{8}{l}{Isotropic decay} \\
HIA & \rev{5.54--13.31} & \rev{5.55--13.32} & -- & -- & -- & -- & -- \\
HIC & \rev{5.49--13.13} & \rev{5.45--13.14} & -- & -- & -- & -- & -- \\
HID & \rev{5.05--13.18} & \rev{5.46--13.20} & -- & -- & -- & -- & -- \\
\addlinespace[2pt]
\multicolumn{8}{l}{Plane strain} \\
PXA & \rev{8.07--10.72} & \rev{8.00--10.75} & 4.00 & 7.02/6.66 & 4.00 & 0.49/0.50 & 1.57 \\
PXB & \rev{8.35--10.81} & \rev{8.29--10.97} & 4.00 & 6.71/6.76 & 4.00 & 0.83/0.85 & 2.95 \\
PXC & \rev{8.50--10.42} & \rev{8.45--10.61} & 4.01 & 6.43/6.61 & 4.01 & 1.19/1.23 & 3.09 \\
PXD & \rev{8.54--9.82} & \rev{8.50--10.02} & 4.01 & 6.20/6.39 & 4.01 & 1.52/1.57 & 3.30 \\
PXE & \rev{8.33--8.82} & \rev{8.39--8.77} & 4.04 & 5.98/6.10 & 4.04 & 1.96/2.03 & 3.34 \\
PXF & \rev{8.06--8.82} & \rev{8.21--8.75} & 3.89 & 5.78/5.88 & 3.88 & 2.05/2.08 & 1.92 \\
PXG & \rev{7.92--8.82} & \rev{8.08--8.75} & 3.89 & 5.71/5.87 & 3.80 & 2.06/2.08 & 1.23 \\
PXH & \rev{7.62--10.08} & \rev{7.53--10.02} & 4.01 & 7.77/6.59 & 4.01 & 0.26/0.25 & 2.22 \\
\addlinespace[2pt]
\multicolumn{8}{l}{Axisymmetric contraction} \\
AXK & \rev{8.95--10.10} & \rev{8.98--10.11} & 3.33 & 5.33/5.08 & 3.33 & 0.40/0.39 & 1.81 \\
AXL & \rev{9.52--10.10} & \rev{9.64--10.11} & 3.36 & 6.57/6.50 & 3.37 & 1.53/1.54 & 1.43 \\
AXM & \rev{9.06--10.10} & \rev{9.17--10.11} & 3.36 & 6.77/6.72 & 3.38 & 2.03/2.05 & 0.77 \\
\addlinespace[2pt]
\multicolumn{8}{l}{Axisymmetric expansion} \\
EXO & \rev{8.63--12.14} & \rev{8.65--11.93} & 3.32 & 5.62/5.31 & 3.32 & 0.35/0.34 & 2.38 \\
EXP & \rev{9.57--11.85} & \rev{9.61--11.81} & 3.34 & 2.99/2.98 & 3.34 & 1.20/1.19 & 1.02 \\
EXQ & \rev{9.57--9.85} & \rev{9.59--9.89} & 3.09 & 1.90/1.88 & 3.09 & 1.45/1.44 & 1.88 \\
\bottomrule
\end{tabular}
\caption{Comparison with Lee--Reynolds benchmarks. \rev{Reference $Re_\lambda$
ranges are converted from the tabulated $Re_\ell$ ranges.}
Anisotropy references are taken from the published tables. $\dagger$
Energy references are digitized at $c_E$. Present energy and anisotropy
values use the nearest saved record to each reference strain. Maximum
energy errors use all sampled strains.}
\label{tab:lee_reynolds_main}
\end{table}

\rev{Table~\ref{tab:lee_reynolds_main} uses the Taylor microscale Reynolds
number defined in Eq.~\eqref{eq:taylor_reynolds_definition}.
\citet{leeReynolds1985} use the velocity scale $q=\sqrt{2K}$, giving
$\sqrt{5Re_\ell}$ for their Taylor microscale Reynolds number. With the
same Taylor microscale and our velocity scale $u'=q/\sqrt{3}$, the
conversion is $Re_\lambda=\sqrt{5Re_\ell/3}$. We apply this conversion to
their $Re_\ell$ ranges in Tables~5.1--5.3 and B.1.}

\rev{For these constant, symmetric mean velocity gradients,
$S_{\mathrm{eq}}$ is evaluated from Eq.~\eqref{eq:rapid_distortion_parameter}
with $S_{ij}=A_{ij}$.} The accumulated strain is expressed as
\begin{equation}
 c=\exp[S_{\mathrm{eq}}(t-t_0)].
 \label{eq:accumulated_strain}
\end{equation}
We compare anisotropy at the tabulated strain $c_b$ and energy at the
digitized strain $c_E$.

Across the fourteen primary deformations, the largest pointwise
relative difference in normalized energy is 3.34\% (PXE).
The largest relative difference in the tabulated anisotropy invariant
occurs for PXH: at $c=4.007$, $-II_b$ is $15.1\%$ below the reference,
while its maximum normalized-energy difference is $2.22\%$.

\citet{zusiPerot2013} showed that the anisotropy of plane-strained
turbulence is sensitive to the initial realization. We therefore repeat
PXH with five random seeds and otherwise identical settings.
The mean $L_{\mathrm{int}}$ grows from $0.3048\pm0.0017$ to
$0.763\pm0.049$, while the endpoint invariants at $c=4.007$ are
$-II_b=0.0698\pm0.0082$ and $10^3III_b=1.9\pm1.5$
(mean $\pm$ sample standard deviation).
Despite comparable integral scale growth, the final anisotropy remains
sensitive to the initial realization. Both reference invariants in
\citet[Table~6.1, H4R]{leeReynolds1985} lie within the observed ranges.

\begin{figure}[p]
\centering
\includegraphics[width=\linewidth]{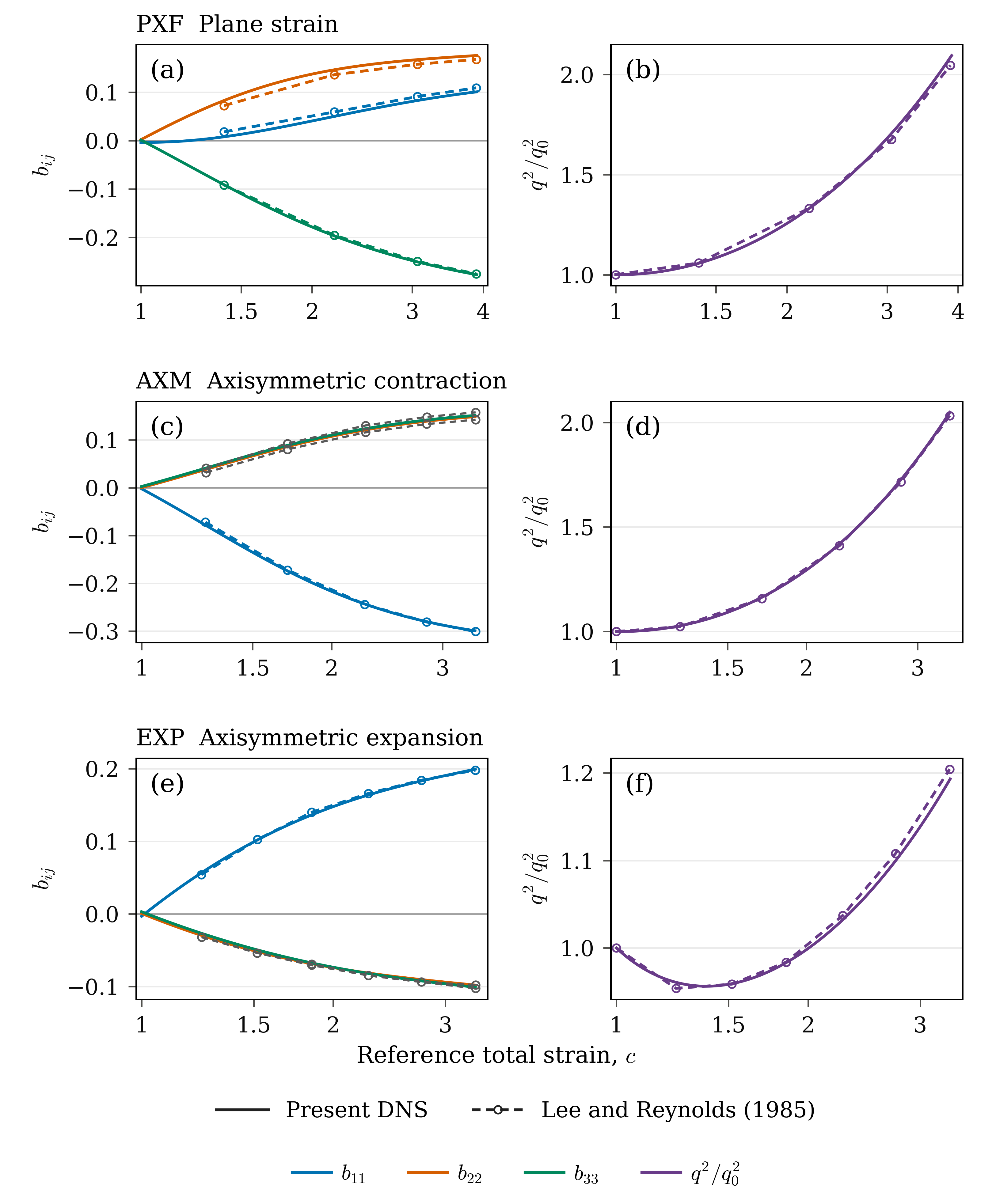}
\caption{\rev{Evolution of the diagonal anisotropy components (left) and
normalized kinetic energy $q^2/q_0^2=K/K_0$ (right) with reference total
strain $c$ for PXF, AXM and EXP. Solid lines show the present DNS. Dashed
lines connect the reference DNS samples shown in Figs.~5.3--5.4, 5.7--5.8
and 5.11--5.12 of \citet{leeReynolds1985}.}}
\label{fig:lee_reynolds_histories}
\end{figure}

\rev{Figure~\ref{fig:lee_reynolds_histories} shows close agreement in the
evolution of the diagonal Reynolds stress anisotropy components and normalized
kinetic energy for one case from each deformation family.}

\subsection{Blaisdell--Shariff benchmarks}

We further validate the solver using the shear and elliptic streamline
cases of \citet{blaisdellShariff1996}. We follow their case settings and
initial-field preparation, then advance the flows with the present RK4/CF24
algorithm and lattice remeshing. \rev{The imposed mean velocity gradient is}
\begin{equation}
 \rev{\mathbf A=
 \begin{pmatrix}
 0 & 0 & -r-e\\
 0 & 0 & 0\\
 r-e & 0 & 0
 \end{pmatrix},}
 \label{eq:blaisdell_mean_gradient}
\end{equation}
\rev{where $e>0$ is the strain rate and $r$ is the mean rotation rate.
The S1 shear case has $r=e$. For the elliptic cases E2--E5, the ratios
$r/e$ are $41/9$, $13/5$, $5/3$, and $5/4$, respectively.
The product $et$ is the nondimensional time, with $t$ measured from the start
of the imposed deformation.}
We compare the generalized skewness $S$
defined in Eq.~\eqref{eq:generalized_derivative_skewness}, the normalized
kinetic energy $K/K_0$ for E2 and E5, and the normalized energy growth rate
\begin{equation}
 \gamma=\frac{P-\epsilon}{eK}.
 \label{eq:normalized_energy_growth_rate}
\end{equation}

\begin{figure}[htbp]
\centering
\includegraphics[width=\linewidth]{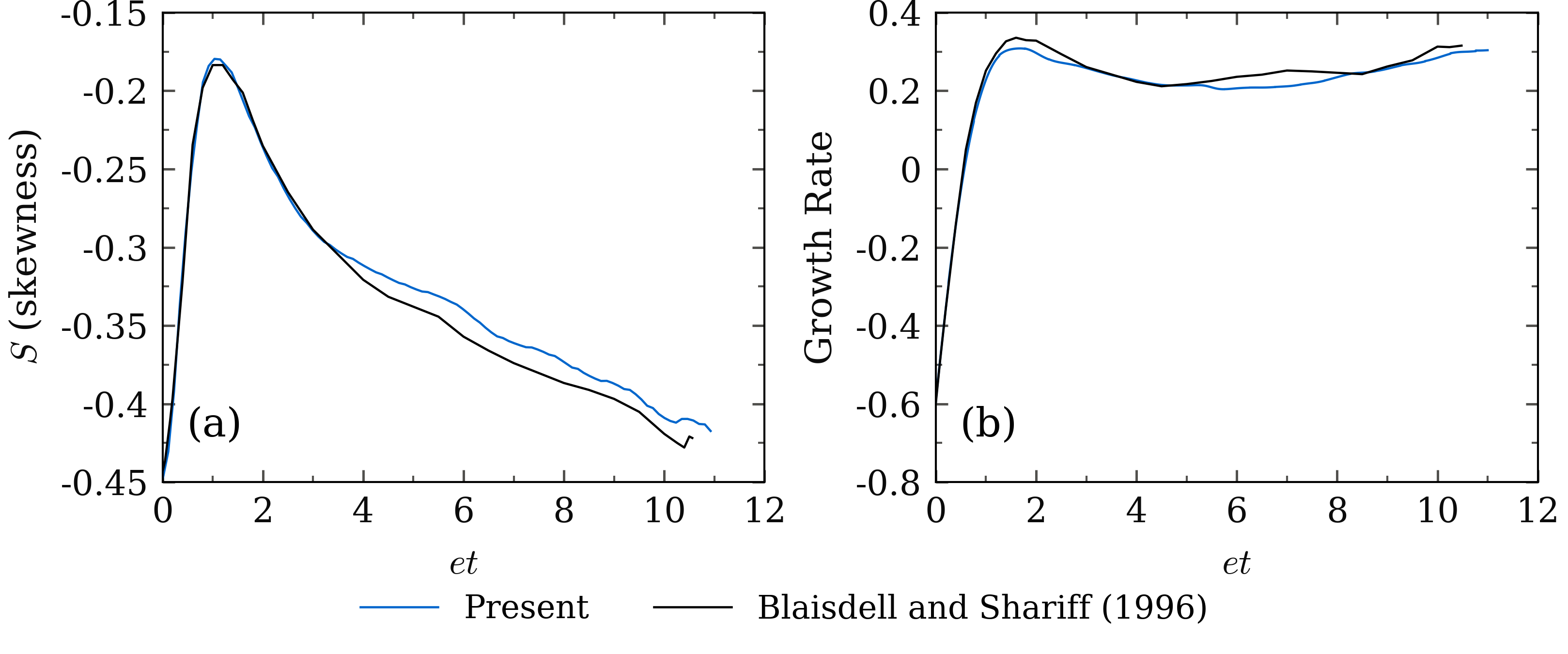}
\caption{Blaisdell--Shariff S1 benchmark through $et=11$: (a) generalized
derivative skewness and (b) energy growth rate. Thin blue lines show the
present DNS; black lines show reference curves digitized from the
published Fig.~5. \rev{Small jumps at remeshes might not be captured at the
sampling frequencies used.}}
\label{fig:blaisdell_s1_response}
\end{figure}

Figure~\ref{fig:blaisdell_s1_response} compares S1 skewness and energy
growth rate with the published reference curves. The initial rise and
subsequent decrease in skewness, together with the positive energy growth
following the initial transient, agree closely with the reference.
The maximum absolute differences from the digitized reference curves
are $0.0196$ in $S$ and $0.0557$ in $\gamma$.

All five cases accept a remesh after the first timestep, using the same
signed permutation $\mathbf M_{\mathrm{first}}$ of the second and third lattice
vectors. This increases the physical cutoff while leaving
$\operatorname{cond}_2(\mathbf B)$ unchanged. During the subsequent shear
evolution in S1, the general remeshing search selects standard integer
shear updates $\mathbf M$, with

\begin{equation}
 \mathbf M_{\mathrm{first}}=
 \begin{pmatrix}
 1&0&0\\0&0&-1\\0&1&0
 \end{pmatrix},
 \qquad
 \mathbf M=\begin{pmatrix}
 1&1&0\\0&1&0\\0&0&1
 \end{pmatrix}.
 \label{eq:s1_integer_shear_remesh}
\end{equation}

\begin{figure}[htbp]
\centering
\includegraphics[width=\linewidth]{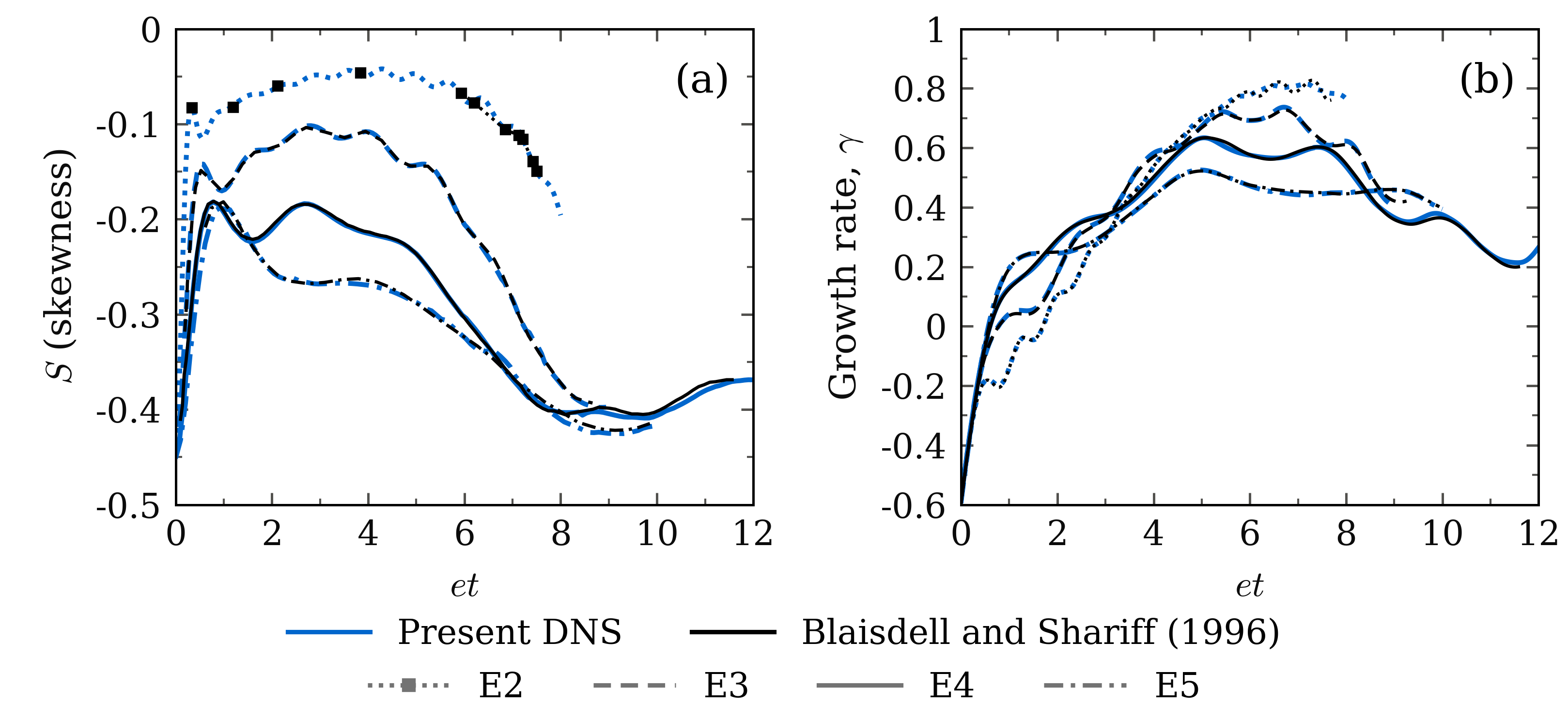}
\caption{\rev{Elliptic flow comparisons for E2--E5. (a) Generalized derivative
skewness. (b) Normalized kinetic energy growth rate. Blue curves show the
present DNS through $et=8,9,12,10$ for E2--E5, respectively. Black curves show
the reference DNS of \citet{blaisdellShariff1996}. Reference skewness is
digitized from their Fig.~4(a), with markers at the identifiable E2 samples.
Reference growth rates are computed from their archived statistics using
Eq.~\eqref{eq:normalized_energy_growth_rate}.}}
\label{fig:blaisdell_response}
\end{figure}

\begin{figure}[htbp]
\centering
\includegraphics[width=\linewidth]{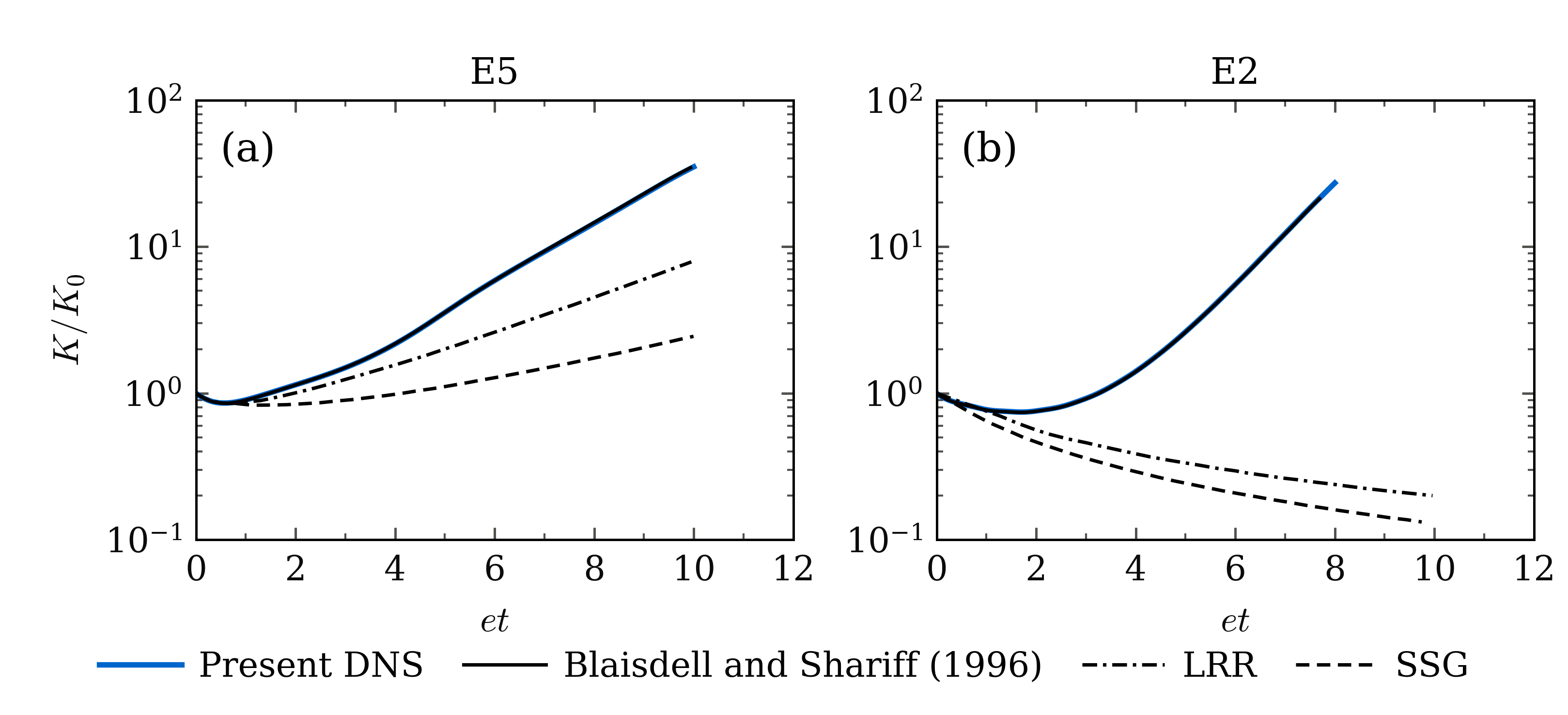}
\caption{\rev{Normalized kinetic energy for (a) E5 and (b) E2. Blue lines
show the present DNS through $et=10$ and $et=8$, respectively. Black solid
lines show the archived DNS of \citet{blaisdellShariff1996}. Black dash-dotted
lines show the Launder--Reece--Rodi (LRR) model and black dashed lines show the
Speziale--Sarkar--Gatski (SSG) model, digitized from their Fig.~7.}}
\label{fig:blaisdell_energy}
\end{figure}

Our solver reproduces the weakly negative skewness
of E2 and the stronger negative skewness developing in E3--E5
(Fig.~\ref{fig:blaisdell_response}). The kinetic energy also follows
the reference transition from decay to amplification
(Fig.~\ref{fig:blaisdell_energy}).

Interestingly, the initial permutation moves the third lattice vector into the
second position, so the subsequent remeshes in E3--E5 take the form of integer
shear updates of either sign in computational coordinates:
\begin{equation}
 \mathbf M_{\pm}=
 \begin{pmatrix}
 1&\pm1&0\\0&1&0\\0&0&1
 \end{pmatrix}.
 \label{eq:elliptic_remesh_matrices}
\end{equation}

\begin{figure}[htbp]
\centering
\includegraphics[width=\linewidth]{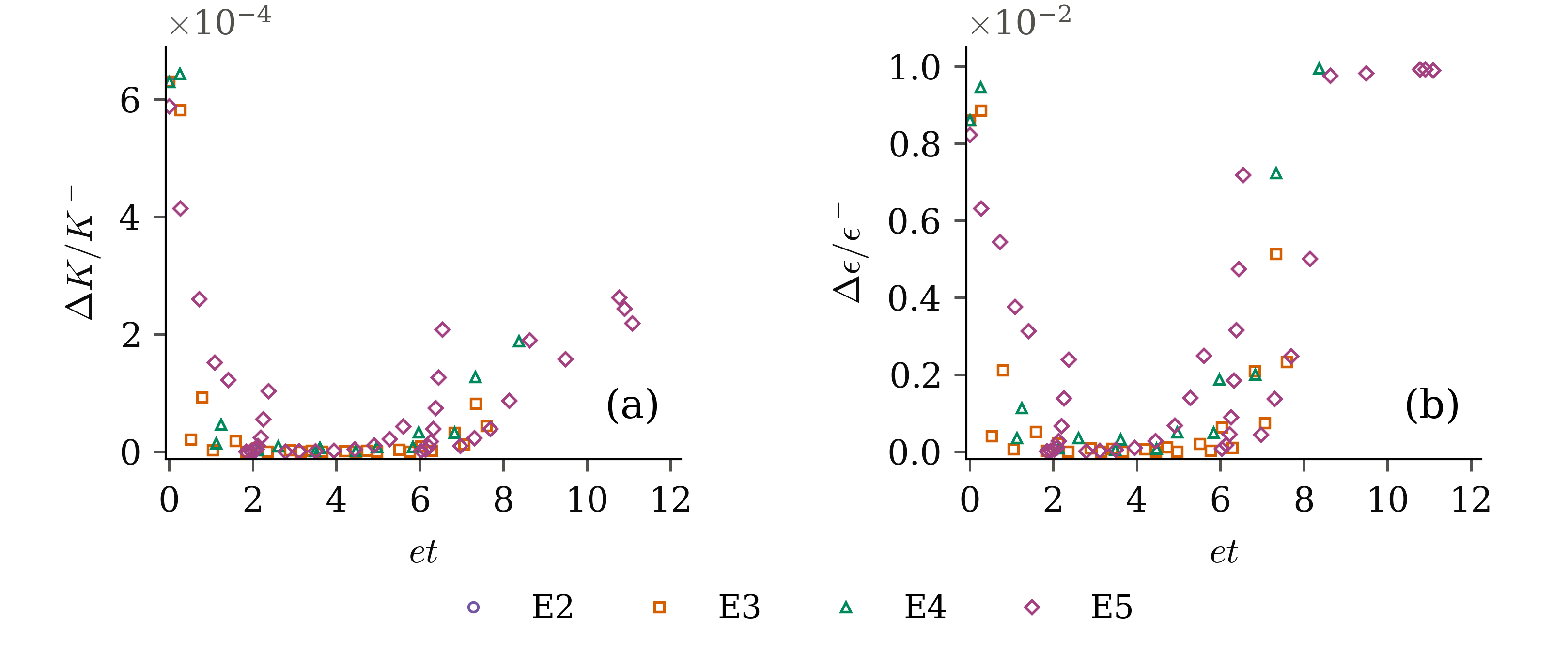}
\caption{Remeshing audit for E2--E5 through $et=8$, $10$, $12$, and $12$,
respectively: (a)
fractional kinetic energy loss, $\Delta K/K^-$, and (b) fractional
dissipation loss, $\Delta\epsilon/\epsilon^-$, at each accepted remesh.}
\label{fig:blaisdell_remeshing}
\end{figure}

Figure~\ref{fig:blaisdell_remeshing} covers the full trajectories,
including portions beyond the benchmark comparison intervals.
Cases E2--E5 undergo $1$, $25$, $15$, and $38$ remeshes, respectively.
Each remesh removes less than $0.065\%$ of the instantaneous
kinetic energy and $1\%$ of the dissipation, within the prescribed
loss limits. In E5, the cumulative loss is $5.08\%$ of $K_0$.
Over the same run, $K$ grows to
$64.55K_0$ and $Re_\lambda$ rises from $9.22$ to $101.07$, so the
discarded energy is $0.079\%$ of the final kinetic energy. Additionally,
the maximum recorded divergence remains below $2.9\times10^{-16}$.

\subsection{\texorpdfstring{\rev{Summary of benchmark cases}}{Summary of benchmark cases}}

\begin{figure}[!htbp]
\centering
\includegraphics[width=0.85\linewidth]{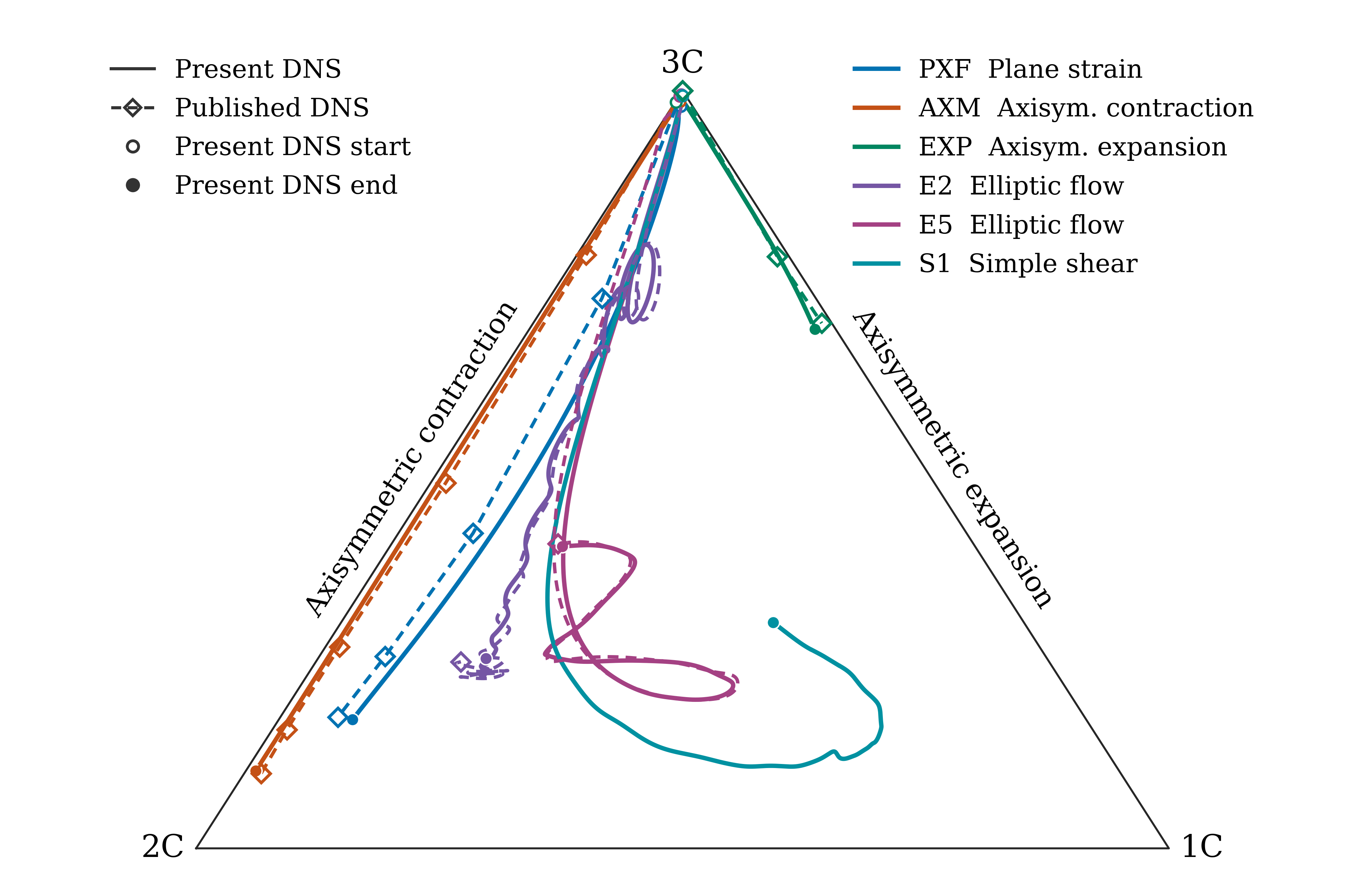}
\caption{\rev{Reynolds stress anisotropy for the Lee--Reynolds, elliptic flow
and simple shear benchmarks. Solid lines show the present DNS with
remeshing. Open and filled circles mark the start and end of the present DNS,
respectively. For S1, only the present DNS is shown, through $et=12$.
Dashed lines show the Lee--Reynolds reference data \citep{leeReynolds1985}
and the E2 and E5 histories from the
\href{https://engineering.purdue.edu/AAE/research/Groups/blaisdell/Elliptic-Flows-and-Strained-Vortices}{authors' data archive}
(\texttt{ctr1996stats.tar}) \citep{blaisdellShariff1996}.
Diamonds mark the Lee--Reynolds samples, reconstructed from tabulated
invariants or digitized diagonal components, and the E2 and E5 reference
endpoints.}}
\label{fig:barycentric_benchmarks}
\end{figure}

\rev{We use the barycentric map of \citet{banerjeeErtuncDurst2008} to compare the
evolution of Reynolds stress anisotropy in the benchmark cases
(Fig.~\ref{fig:barycentric_benchmarks}). AXM and EXP remain close to the
contraction and expansion edges, respectively. The S1 trajectory approaches
the two-component edge before turning back toward the interior, forming a
hook.}

\rev{In E5, the opposite signs of $A_{13}$ and $A_{31}$ in
Eq.~\eqref{eq:blaisdell_mean_gradient} make the box deformation periodic.
The unremeshed box shears and stretches in the $x_1$--$x_3$ plane and
recovers its original shape every $4\pi/3$ in $et$. We reconstruct the
physical cutoff for the unremeshed DNS of \citet{blaisdellShariff1996} from
$\mathbf B=\exp(-\mathbf A t)$ and its reported grid and Fourier truncation
using Eq.~\eqref{eq:physical_resolution}.
Near maximum distortion at $et\approx2.09$ and $6.28$, the remeshed cutoff
is twice the reference cutoff. At the last common time, $et=9.94$,
the physical $k_{\max}\eta$ is $0.56$ in the remeshed run and $0.22$ in
the reference run, using the dissipation from each run to calculate
$\eta$. The maximum absolute difference in the normalized shear stress
$b_{13}$ is $0.0022$ over the common interval.}

\section{General mean deformation cases}
\label{sec:general_cases}

\setlength{\textfloatsep}{12pt plus 2pt minus 2pt}
\setlength{\floatsep}{10pt plus 2pt minus 2pt}
\setlength{\intextsep}{10pt plus 2pt minus 2pt}

The General case combines strain, shear, and rotation with changing principal
directions. The General Shear and Rapid General Shear cases use constant three-dimensional shear at
different rates.
The runs are identified in Table~\ref{tab:general_cases}.
The General and General Shear cases use the same selected field at each Reynolds number, and
the Rapid General Shear case uses the $Re_\lambda=33.211$ field
(Table~\ref{tab:initial_fields}).
Paired runs differ only in whether remeshing is enabled.

\begin{table}[htbp]
\centering
\small
\setlength{\tabcolsep}{3pt}
\begin{tabular}{@{}llccc@{}}
\toprule
Case ID & Flow & $Re_{\lambda,0}$ & Grid & $\widehat t_f$ \\
\midrule
\multicolumn{5}{@{}l}{General (G)} \\
G-15 & General mean rotation and strain & 15.207 & $128^3$ & 8 \\
G-15-fine & General mean rotation and strain & 15.207 & $256^3$ & 8 \\
G-33 & General mean rotation and strain & 33.211 & $256^3$ & 8 \\
\addlinespace[2pt]
\multicolumn{5}{@{}l}{General Shear (GS)} \\
GS-15 & General shear with mean rotation and strain & 15.207 & $128^3$ & 6 \\
GS-33 & General shear with mean rotation and strain & 33.211 & $256^3$ & 6 \\
\addlinespace[2pt]
\multicolumn{5}{@{}l}{Rapid General Shear (RGS)} \\
RGS-33 & Rapid general shear with mean rotation and strain & 33.211 & $256^3$ & 0.06 \\
\bottomrule
\end{tabular}
\caption{General mean deformation cases with final times $\widehat t_f=t_f/T_r$.
\rev{The prescribed mean velocity gradients are defined in Section~\ref{sec:general_histories}.}
The suffixes $15$ and $33$ identify the initial Reynolds numbers.
Each case is simulated with and without remeshing, except for the fine-grid
simulation, which uses remeshing. The Rapid General Shear case also includes viscous RDT
simulations with and without remeshing.}
\label{tab:general_cases}
\end{table}

\subsection{Prescribed flows}
\label{sec:general_histories}

For the General and General Shear cases, we use $T_0=K_0/\epsilon_0$.
In the General case, we set $s=t/(2T_0)$ and combine strain and shear in a rotating frame described by
$\mathbf Q(s)=\mathbf R_z(\omega_zs)\mathbf R_y(\omega_ys)$,
where $\mathbf R_j$ is the rotation matrix about Cartesian axis $j$.
The resulting mean velocity gradient is

\begin{equation}
 \mathbf A(t)=\frac{1}{2T_0}\left[
 \mathbf Q'\mathbf Q^{\mathsf T}
 +\mathbf Q
 \begin{pmatrix}
 0.04&0.6e^{0.05s}&0.24s e^{0.07s}\\
 0&-0.01&0.4e^{0.02s}\\
 0&0&-0.03
 \end{pmatrix}
 \mathbf Q^{\mathsf T}\right],
 \label{eq:general_sheared_lattice}
\end{equation}
where $(\omega_z,\omega_y)=(0.35,0.22)$ and
$\mathbf Q'=d\mathbf Q/ds$. The first term rotates the prescribed
frame. The second gives strain and shear in that frame.
The gradient is trace free and generally satisfies
$\mathbf A(t_1)\mathbf A(t_2)\ne\mathbf A(t_2)\mathbf A(t_1)$.
This case is advanced over $0\le\widehat t\le8$.

The General Shear case combines mean strain and rotation in a
three-dimensional shear flow with constant velocity gradient
\begin{equation}
 \mathbf A_{\mathrm{GS}}=\frac{1}{9T_0}
 \begin{pmatrix}
 2&5&-4\\
 2&-4&5\\
 8&2&2
 \end{pmatrix},
 \label{eq:dense_shear_gradient}
\end{equation}
and is compared over $0\le\widehat t\le6$.
The Rapid General Shear case uses $\mathbf A_{\mathrm{RGS}}=c\mathbf A_{\mathrm{GS}}$, with $c=100$ and
$Re_{\lambda,0}=33.211$, and the comparison extends to $\widehat t=0.06$.
The constant gradients are trace free. The elapsed physical time
in the Rapid General Shear case is smaller than in the General Shear case by a factor of $100$.

\subsection{Summary of the simulations}
\label{sec:general_response}

\begin{figure}[!htbp]
\centering
\includegraphics[width=\linewidth]{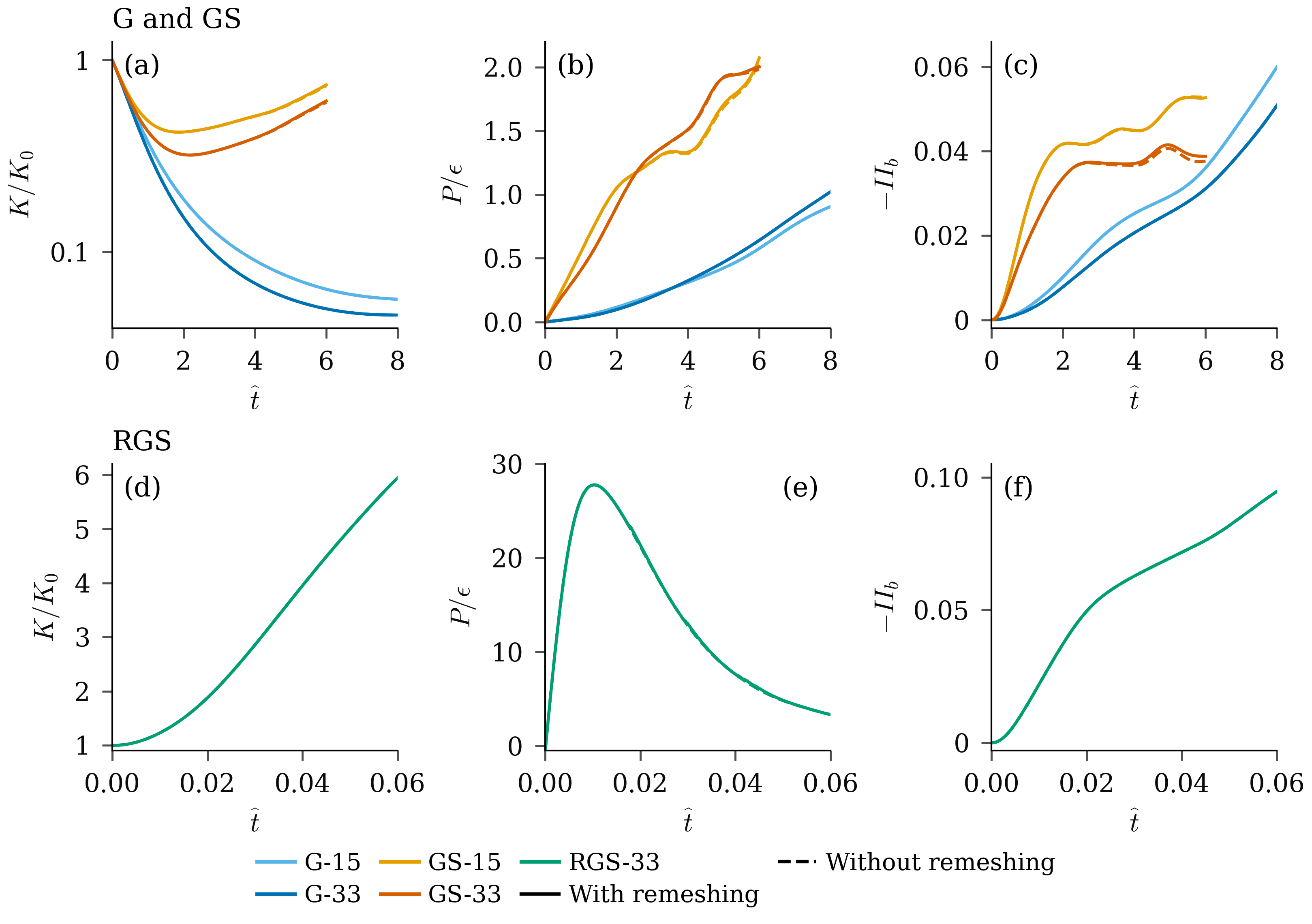}
\caption{Turbulence response for the General and General Shear cases at
both initial Reynolds numbers (top row) and the Rapid General Shear case
at $Re_{\lambda,0}=33.211$
(bottom row). Columns show normalized kinetic energy, production relative
to dissipation, and the anisotropy invariant $-II_b$.}
\label{fig:general_response}
\end{figure}

Figure~\ref{fig:general_response} shows strong energy decay in
the General case, decay followed by recovery in the General Shear case, and strong amplification
in the Rapid General Shear case. The faster deformation in the Rapid General Shear case increases production
relative to dissipation. The bulk statistics with and without
remeshing remain close in all three cases.
Table~\ref{tab:general_endpoints} lists endpoint statistics, resolution
estimates, timestep counts and mean elapsed times per $100$ steps.

\begin{table}[!htbp]
\centering
\footnotesize
\fontsize{9.5}{11.5}\selectfont
\setlength{\tabcolsep}{2pt}
\begin{tabular*}{\linewidth}{@{\extracolsep{\fill}}lcc@{\hspace{6pt}}rrrrcrr@{}}
\toprule
& & & & & \multicolumn{2}{c}{$k_{\max}\eta$} & & & \\
\cmidrule(lr){6-7}
Case & Remeshing & Grid & $K_f/K_0$ & $-II_{b,f}$ & Physical & \shortstack{Compu-\\tational}
& \shortstack{Ranks/\\threads} & Steps & \shortstack{s/100\\steps} \\
\midrule
G-15 & Yes & $128^3$ & 0.057 & 0.060 & 2.75 & 3.74 & 32 & 1958 & 127.9 \\
G-15 & No & $128^3$ & 0.057 & 0.060 & 0.65 & 3.74 & 32 & 1958 & 126.5 \\
G-15-fine & Yes & $256^3$ & 0.057 & 0.060 & 5.49 & 7.49 & 32 & 1998 & 1153.6 \\
G-33 & Yes & $256^3$ & 0.047 & 0.051 & 2.78 & 3.79 & 32 & 2162 & 1158.3 \\
G-33 & No & $256^3$ & 0.047 & 0.051 & 0.66 & 3.79 & 32 & 2171 & 1158.0 \\
\addlinespace[2pt]
GS-15 & Yes & $128^3$ & 0.74 & 0.053 & 1.13 & 1.95 & 32 & 2610 & 136.7 \\
GS-15 & No & $128^3$ & 0.74 & 0.053 & 0.098 & 1.96 & 32 & 4821 & 142.1 \\
GS-33 & Yes & $256^3$ & 0.61 & 0.039 & 1.11 & 1.92 & 32 & 2609 & 1163.9 \\
GS-33 & No & $256^3$ & 0.60 & 0.038 & 0.097 & 1.94 & 32 & 11331 & 1237.6 \\
\addlinespace[2pt]
RGS-33 & Yes & $256^3$ & 5.94 & 0.095 & 0.098 & 0.48 & 32 & 2609 & 1212.6 \\
RGS-33 & No & $256^3$ & 5.94 & 0.095 & 0.024 & 0.48 & 32 & 2609 & 1163.1 \\
\addlinespace[2pt]
RGS-33-RDT & Yes & $256^3$ & 5.91 & 0.098 & 0.14 & 0.48 & 8 & 2609 & -- \\
RGS-33-RDT & No & $256^3$ & 5.91 & 0.098 & 0.024 & 0.48 & 8 & 2609 & -- \\
\bottomrule
\end{tabular*}
\caption{DNS and RDT statistics and computational cost. The comparison times are
$\widehat t=8$ for General, $6$ for General Shear, and $0.06$ for Rapid General Shear.
RGS-33-RDT denotes viscous RDT for RGS-33.
DNS uses $32$ \rev{Message Passing Interface (MPI)} ranks, and RDT uses $8$
local threads.}
\label{tab:general_endpoints}
\end{table}

\begin{figure}[!htbp]
\centering
\includegraphics[width=0.85\linewidth]{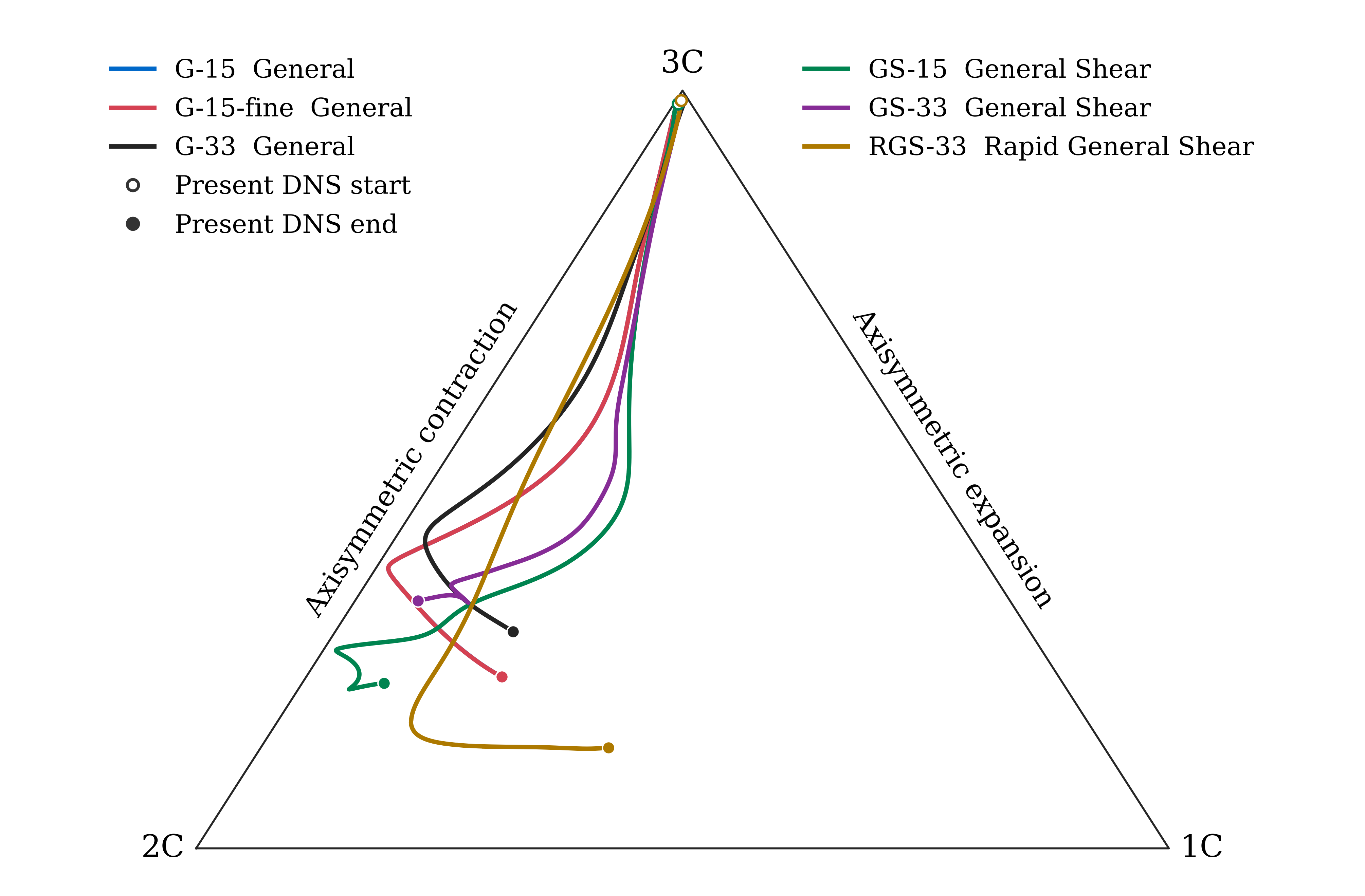}
\caption{\rev{Reynolds stress anisotropy for the general mean deformation
cases with remeshing. Open and filled circles mark the start and end of the
present DNS, respectively.
The G-15 trajectories on the $128^3$ and $256^3$ grids nearly coincide.}}
\label{fig:barycentric_general}
\end{figure}

\rev{Figure~\ref{fig:barycentric_general} compares the evolution of Reynolds
stress anisotropy in the General, General Shear and Rapid General Shear
cases. The trajectories show different paths through the interior of the
barycentric map, with Rapid General Shear reaching closest to the
two-component edge.}

\subsection{Remeshing analysis}
\label{sec:remeshing_needed}

\begin{figure}[!htbp]
\centering
\includegraphics[width=0.96\linewidth]{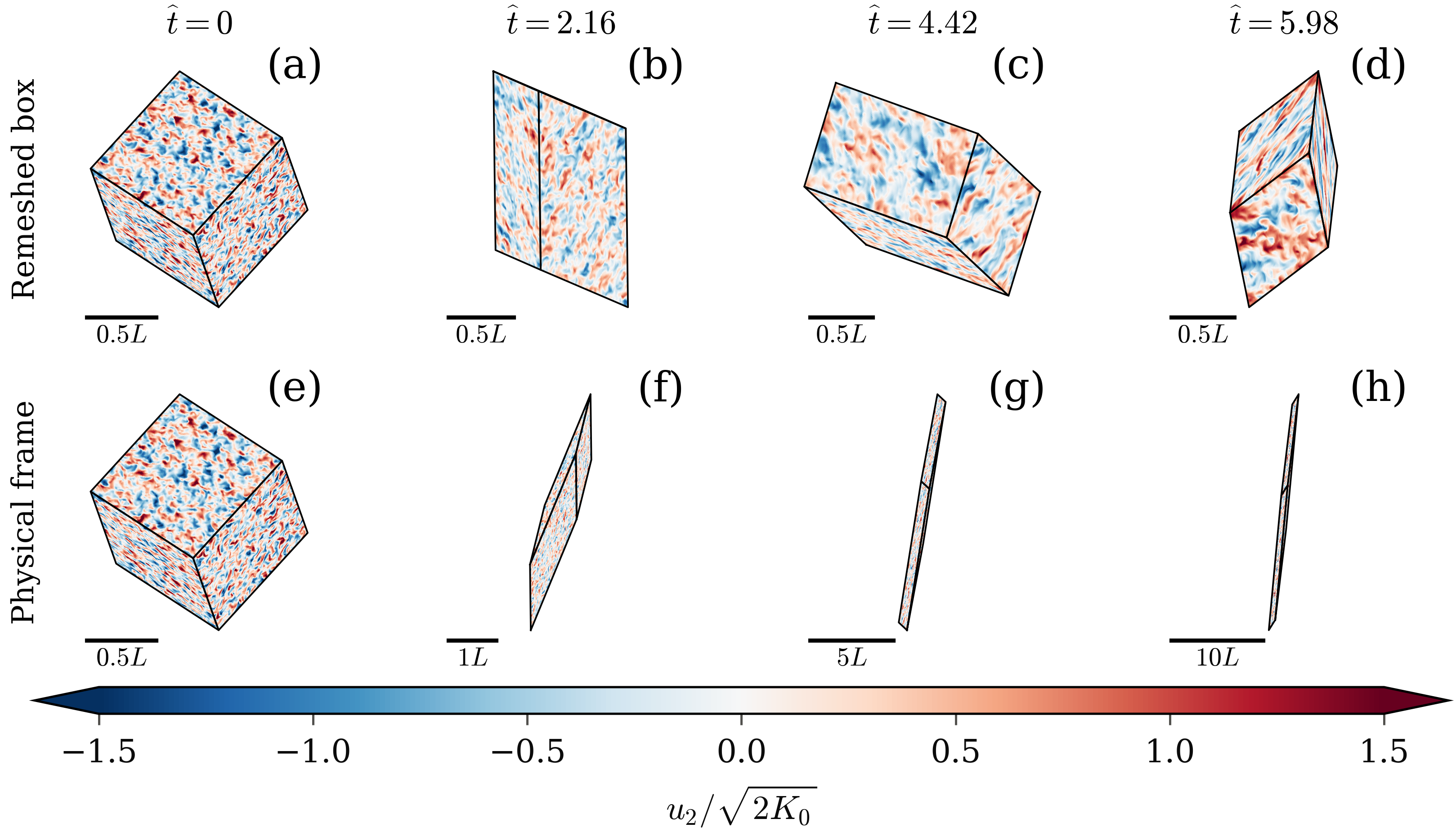}
\caption{The General Shear case at $Re_{\lambda,0}=33.211$ on the $256^3$ grid. Columns
show $\widehat t=0$, $2.16$, $4.42$, and $5.98$. The middle columns show the
states immediately after the sixth and twelfth remeshes. The upper row shows the periodic
box used after remeshing. The lower row shows the continuously deformed
box in the physical frame. \rev{Colours show $u_2/\sqrt{2K_0}$, the velocity
fluctuation along the physical $x_2$ axis normalized by the initial root-mean-square
fluctuation speed.} The viewing direction is fixed. \rev{Each panel is scaled to fit,
and all scale bars use the same initial box side $L$.}}
\label{fig:general_evolution}
\end{figure}

As discussed in Section~\ref{sec:remeshing}, the map
$\boldsymbol\kappa=\mathbf B^{\mathsf T}\mathbf k$ changes the physical
wavenumber range resolved in each direction. Remeshing changes the
lattice basis to improve coverage in poorly resolved directions and reduce
distortion of the computational box (Fig.~\ref{fig:general_evolution}).

In the General Shear run at $Re_{\lambda,0}=33.211$ on the $256^3$ grid
illustrated in Fig.~\ref{fig:general_evolution}, the sixth and twelfth remeshes, at
$\widehat t=2.164$ and $4.418$, use the basis changes

\begin{equation}
 \mathbf M_6=\begin{pmatrix}
 1&1&0\\-1&0&0\\1&0&1
 \end{pmatrix},
 \qquad
 \mathbf M_{12}=\begin{pmatrix}
 1&0&0\\-1&1&0\\0&0&1
 \end{pmatrix}.
 \label{eq:general_snapshot_remeshes}
\end{equation}

$\mathbf M_6$ replaces $\mathbf h_1$ by
$\mathbf h_1-\mathbf h_2+\mathbf h_3$ and $\mathbf h_2$ by $\mathbf h_1$.
Its relative energy and dissipation losses are
$\Delta K/K^-=3.41\times10^{-5}$ and
$\Delta\epsilon/\epsilon^-=1.63\times10^{-3}$.
$\mathbf M_{12}$ replaces $\mathbf h_1$ by $\mathbf h_1-\mathbf h_2$.
Its relative losses are $1.41\times10^{-5}$ and $1.22\times10^{-3}$.
The physical cutoff increases by $4.40\%$ and $5.86\%$, respectively.

To assess the effect of remeshing beyond kinetic energy, we examine the
dissipation tensor
$\epsilon_{ij}=2\nu\langle(\partial u_i/\partial x_k)
(\partial u_j/\partial x_k)\rangle$
and the slow pressure rate of strain tensor $\mathcal R^{(s)}_{ij}$.
Both appear as unclosed terms in turbulence transport equations
\citep{pope2000}.

Without remeshing, the physical $k_{\max}\eta$ decreases to about $0.65$
in the General case, where the compared statistics remain nearly unchanged.
In the General Shear case, it falls by more than a factor of twelve from its initial
value to about $0.10$, whereas remeshing maintains endpoint values near
$1.1$ (Table~\ref{tab:general_endpoints}). This greater loss of physical
resolution is consistent with the larger differences in shear stress and
its budget terms (Table~\ref{tab:remeshing_comparison}).

\begin{table}[!htbp]
\centering
\small
\setlength{\tabcolsep}{3pt}
\begin{tabular}{lrrrrrrr}
\toprule
& & \multicolumn{4}{c}{Change due to remeshing (\%)} &
\multicolumn{2}{c}{\rev{Summed losses (\%)}} \\
\cmidrule(lr){3-6}\cmidrule(lr){7-8}
Case & $\widehat t$ & $K$ & $\langle u_2u_3\rangle$ & $\epsilon_{23}$ &
$\mathcal R^{(s)}_{23}$ & \rev{$K$} & \rev{$\epsilon$} \\
\midrule
G-15 & 8 & 0.000341 & 0.000297 & -0.00276 & 0.0434 & \rev{0.00678} & \rev{0.158} \\
G-33 & 8 & 0.00123 & 0.00251 & -0.0291 & 0.0719 & \rev{0.00283} & \rev{0.112} \\
GS-15 & 6 & 1.15 & 23.07 & -8.74 & 16.96 & \rev{0.194} & \rev{4.42} \\
GS-33 & 6 & 1.60 & 23.62 & -66.47 & 40.00 & \rev{0.0802} & \rev{3.03} \\
RGS-33 & 0.06 & 0.000993 & 0.0177 & 0.0238 & 0.809 & \rev{0.140} & \rev{6.07} \\
RGS-33-RDT & 0.06 & -0.0347 & -0.0230 & -0.251 & -- & \rev{0.240} & \rev{7.57} \\
\bottomrule
\end{tabular}
\caption{\rev{This table compares the final statistics of simulations with
and without remeshing and gives the summed losses from discarded Fourier modes.}
The slow pressure term is absent in viscous RDT (RGS-33-RDT).
Each change is computed as $100(X_{\mathrm{with}}/X_{\mathrm{without}}-1)$.
For negative components, a positive change denotes an increase in magnitude.
\rev{The summed losses are $100\sum_r\Delta K_r/K_r^-$ and
$100\sum_r\Delta\epsilon_r/\epsilon_r^-$ over the remeshes $r$.
Equation~\eqref{eq:remesh_truncation_losses} gives the losses in each numerator.
The denominators are the values immediately before each remesh.}}
\label{tab:remeshing_comparison}
\end{table}

\begin{figure}[!htbp]
\centering
\includegraphics[width=\linewidth]{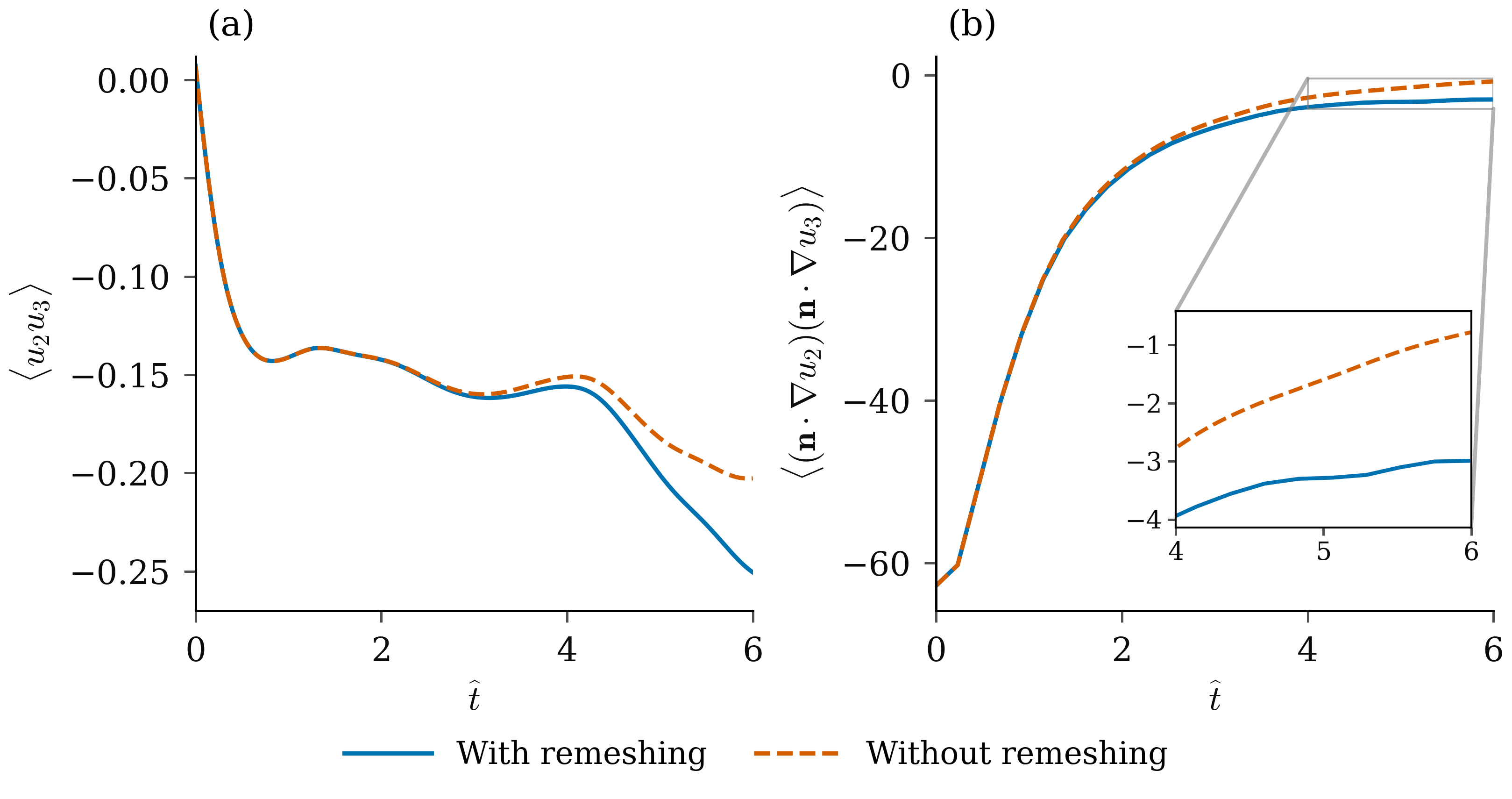}
\caption{The General Shear case at $Re_{\lambda,0}=33.211$ with and
without remeshing: (a) shear stress
$\langle u_2u_3\rangle$ and
(b) directional gradient covariance
$\langle(\mathbf n\cdot\nabla u_2)(\mathbf n\cdot\nabla u_3)\rangle$.
The inset shows $4\le\widehat t\le6$.
The unit vector $\mathbf n\simeq(-0.085,\,0.57,\,0.82)^{\mathsf T}$
gives the direction of minimum physical spectral coverage in the
unremeshed run at $\widehat t=6$.}
\label{fig:remeshing_analysis}
\end{figure}

In the General Shear case at $Re_{\lambda,0}=33.211$, the change
in kinetic energy is only $1.60\%$, while the magnitude of
$\langle u_2u_3\rangle$ changes by $23.62\%$. Kinetic energy contains
only the diagonal Reynolds stresses and therefore does not reveal this
change in shear stress. At the end of the run, the shear stress starts
increasing without remeshing and continues decreasing with remeshing
(Fig.~\ref{fig:remeshing_analysis}).
The directional gradient covariance also remains larger in magnitude with
remeshing [Fig.~\ref{fig:remeshing_analysis}(b)].

\subsection{Rapid distortion comparison}
\label{sec:general_rdt_comparison}

\rev{In both Rapid General Shear DNS runs, $S^*$ from
Eq.~\eqref{eq:rapid_distortion_parameter} starts at $141.4$ and remains above
$22.5$ through $\widehat t=0.06$. We assess the rapid distortion approximation
by comparing DNS with the viscous RDT solution of Eq.~\eqref{eq:viscous_rdt}
at each DNS timestep.}

\begin{figure}[!htbp]
\centering
\includegraphics[width=\linewidth]{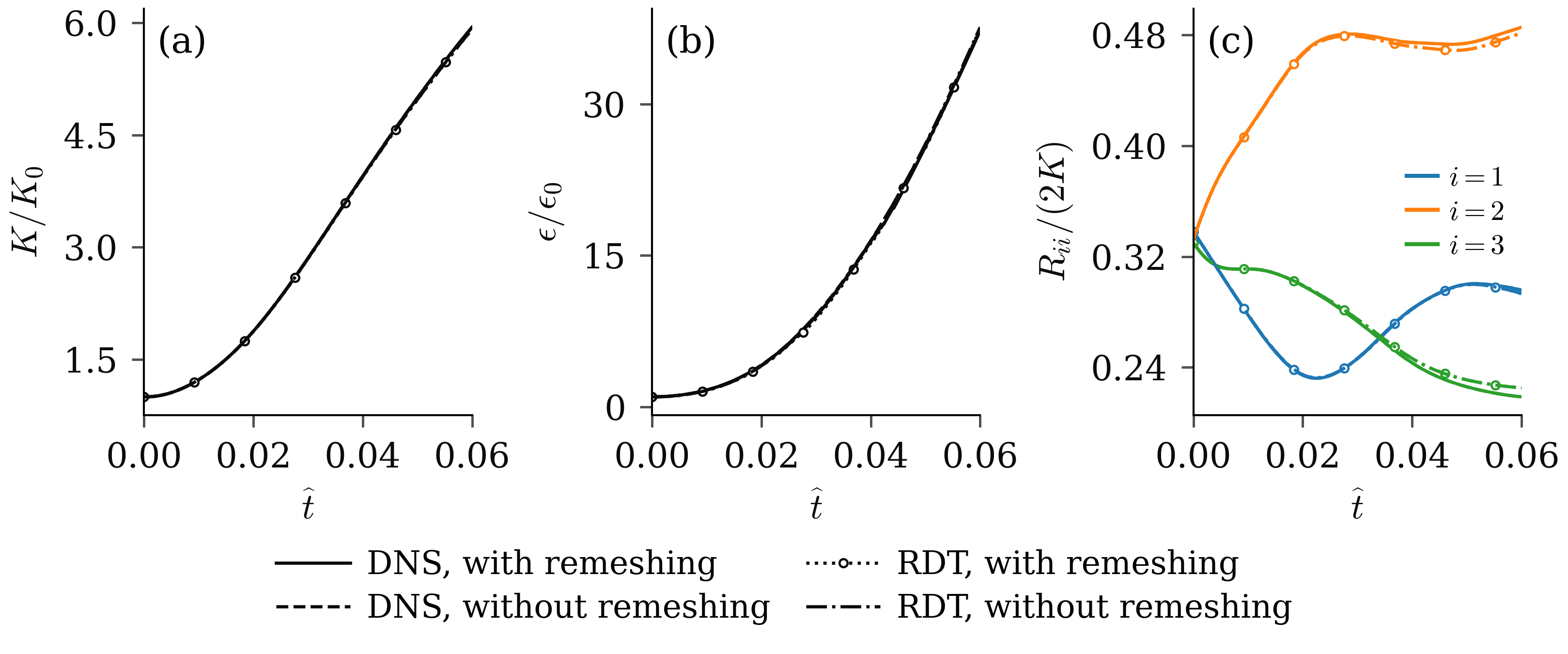}
\caption{Comparison of DNS and viscous RDT for the Rapid General Shear case
with and without remeshing at $Re_{\lambda,0}=33.211$:
(a) kinetic energy $K/K_0$, (b) dissipation $\epsilon/\epsilon_0$, and
(c) component energy fractions $R_{ii}/(2K)$, $i=1,2,3$,
where $R_{ij}=\langle u_i u_j\rangle$ is the Reynolds stress tensor.
Solid and dashed lines denote DNS with and without remeshing.
Dotted lines with open circles and dash-dotted lines denote RDT
with and without remeshing.
Colours distinguish velocity components in (c).}
\label{fig:general_rdt_comparison}
\end{figure}

\rev{For both remeshing settings, DNS and viscous RDT agree closely in
kinetic energy $K$, total dissipation $\epsilon$ and the three component
energy fractions $R_{ii}/(2K)$ (Fig.~\ref{fig:general_rdt_comparison}).}

Despite the reduced directional coverage without remeshing, the shear stress
and $\epsilon_{23}$ are much less sensitive to remeshing in the Rapid General Shear
case than in the General Shear case. \rev{Table~\ref{tab:remeshing_comparison}
compares runs with and without remeshing within each method. At
$\widehat t=0.06$, both quantities change by less than $0.024\%$ in DNS
and less than $0.26\%$ in viscous RDT. The low physical resolution in the
Rapid General Shear case warrants a grid convergence check before using it
as reference data.}

\section{Numerical accuracy and computational cost}
\label{sec:numerical_accuracy}

\subsection{Grid comparison}
\label{sec:grid_comparison}

\begin{figure}[!htbp]
\centering
\includegraphics[width=\linewidth]{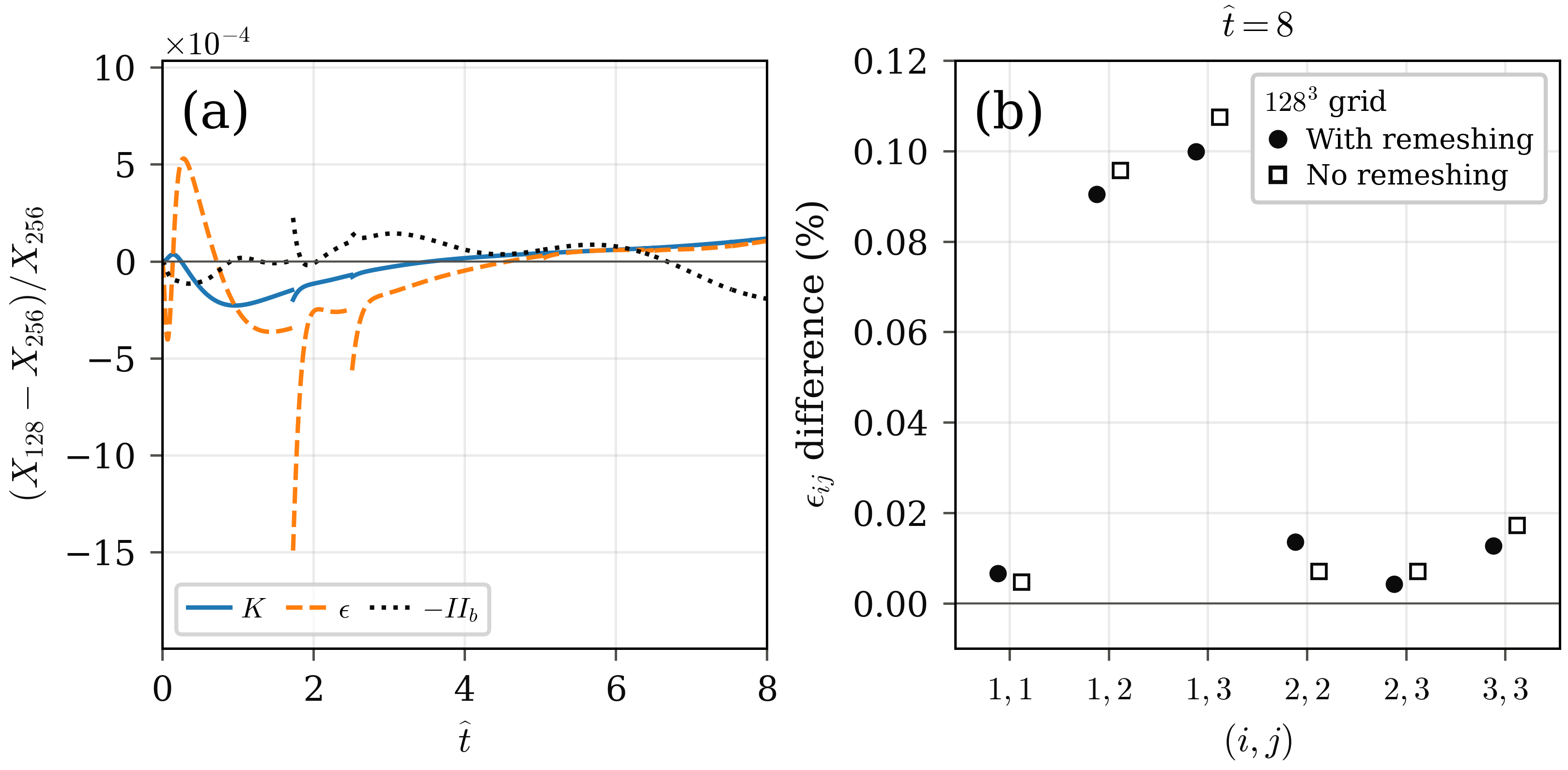}
\caption{Grid comparison for the General case at $Re_{\lambda,0}=15.207$.
(a) Relative differences for $X=K$, $\epsilon$, and $-II_b$ with remeshing.
(b) Percentage differences in the six independent components of
$\epsilon_{ij}$ at $\widehat t=8$, labelled by $(i,j)$.
Both panels use the $256^3$ simulation with
remeshing as the reference.}
\label{fig:general_numerical_accuracy}
\end{figure}

For the General case at $Re_{\lambda,0}=15.207$, the $256^3$ run starts from
the same initial Fourier coefficients as the $128^3$ General run,
transferred by zero-padding without amplitude rescaling.
Figure~\ref{fig:general_numerical_accuracy} compares the $128^3$ runs
with and without remeshing against the $256^3$ reference with remeshing.

Doubling the grid from $128^3$ to $256^3$ changes the final energy,
dissipation, and anisotropy
by less than $0.02\%$ (Fig.~\ref{fig:general_numerical_accuracy}).
At $\widehat t=8$, both $128^3$ runs agree with the $256^3$ reference within
$0.11\%$ in every component of $\epsilon_{ij}$.

\subsection{Computational cost}
\label{sec:computational_cost}

The mean elapsed times per $100$ steps in
Table~\ref{tab:general_endpoints} differ by at most $6.0\%$ between paired
runs. These differences suggest that adding remeshing does not materially
change the average cost per step. Its additional cost appears negligible
compared with the FFTs required at each Runge--Kutta stage.

The General Shear case at $Re_{\lambda,0}=33.211$ takes $11331$
steps without remeshing, compared with $2609$
with remeshing. The CFL condition uses $\mathbf B\mathbf u$
[Eq.~\eqref{eq:computational_cfl}], so the higher advection rate from the
unremeshed coordinate map forces smaller timesteps.
Remeshing reduces the required step count by a factor of $4.34$.
Hence, remeshing reduces the total step count under our CFL condition
without materially increasing the measured time per $100$ steps.

\section{Discussion}
\label{sec:discussion}

\subsection{\texorpdfstring{\rev{Initial Field Forcing and Box effects}}
  {Initial Field Forcing and Box effects}}

Tests with initial fields developed using the $3$--$6$ forcing band
showed growth of $L_{\mathrm{int}}$ relative to the periodic domain during
deformation after forcing was removed. We therefore used the $8$--$16$ band
to start from smaller structures and reduce the risk of box effects 
during deformation. Whether the observed growth reflects
physical evolution, finite-box effects, or both remains unclear.

\subsection{\texorpdfstring{\rev{Importance of Remeshing}}
  {Importance of Remeshing}}

\rev{In the General Shear case at $Re_{\lambda,0}=33.211$, both simulations
use a $256^3$ grid, yet the final physical $k_{\max}\eta$ is $0.10$ without
remeshing and $1.1$ with it (Table~\ref{tab:general_endpoints}). Kinetic
energy and total dissipation differ by $1.6\%$ and $5.4\%$, while the shear
stress and $\epsilon_{23}$ differ by about $24\%$ and $66\%$
(Table~\ref{tab:remeshing_comparison}).
Kinetic energy and total dissipation are half the sums of the diagonal
components of the Reynolds stress and dissipation tensors, respectively.
These sums can remain close between simulations with and without remeshing
even when the shear stress and $\epsilon_{23}$ differ substantially.
At the end of the unremeshed simulation, the least resolved direction is
$\mathbf n\simeq(-0.085,\,0.57,\,0.82)^{\mathsf T}$. Along this direction,
the off-diagonal gradient covariance is substantially larger in magnitude
with remeshing [Fig.~\ref{fig:remeshing_analysis}(b)].}

\rev{In the General cases, every component of the Reynolds stress and
dissipation tensors differs by less than $0.03\%$ between simulations with
and without remeshing at the final time. Despite this, remeshing raises the final physical
$k_{\max}\eta$ from $0.65$ to $2.75$ at $Re_{\lambda,0}=15.207$ and from
$0.66$ to $2.78$ at $Re_{\lambda,0}=33.211$ (Table~\ref{tab:general_endpoints}).
In \citeauthor{blaisdellShariff1996}'s (\citeyear{blaisdellShariff1996}) E5
case, the unremeshed box periodically recovers its original shape. Our
simulation undergoes $38$ remeshes over the full run, and its skewness and
energy growth rate remain close to the unremeshed reference over the
comparison interval (Fig.~\ref{fig:blaisdell_response}). At the last time
shared by our simulation and their data archive, $et=9.94$, the physical
$k_{\max}\eta$ is $0.56$ with remeshing, more than twice the reference value
of $0.22$.}

\subsection{\texorpdfstring{\rev{Truncation losses and alternatives to remeshing}}
  {Truncation losses and alternatives to remeshing}}

\rev{The set of Fourier modes discarded at a remesh depends on the selected
basis, so truncation can affect the subsequent evolution.
In the General Shear cases, discarding modes during remeshing gives
summed dissipation losses of $3.03$--$4.42\%$ and may contribute to the
large differences in shear stress and off-diagonal components of dissipation
tensor. However, the viscous RDT runs show much smaller differences in both quantities
despite a larger summed dissipation loss of $7.57\%$
(Table~\ref{tab:remeshing_comparison}). Similarly, E5 retains close agreement
with the unremeshed reference in skewness and energy growth rate over the
comparison interval, with a summed dissipation loss of about $11\%$ over
the full run.}

\rev{With the current per-remesh limits of $0.1\%$ for kinetic energy and
$1\%$ for dissipation rate, truncation losses could be reduced further by
favoring candidates with smaller predicted losses among those that meet the
spectral coverage and conditioning requirements. \citet{brucker2007} avoided
remeshing in homogeneous simple shear by
evaluating the nonlinear terms on a fixed orthogonal grid using a Fourier
transform adapted to shear. Whether this approach can be extended to maintain
directional spectral coverage under general mean deformation remains an open
question.}

\section{Conclusions}
\label{sec:conclusions}

We extended \citeauthor{rogallo1981}'s (\citeyear{rogallo1981}) pseudospectral
algorithm for DNS of incompressible homogeneous turbulence to general
spatially uniform mean velocity gradients using a full matrix velocity
transformation and updates that preserve unit determinants.
Three-dimensional remeshing uses a bounded search over compositions of
elementary integer column operations to generate candidate bases of the same
physical periodic lattice. Selection prioritizes physical spectral coverage
and conditioning while keeping predicted kinetic energy and dissipation
losses within specified tolerances.
Additionally, the small change in measured average cost per step suggests
that remeshing adds negligible cost compared with the Fourier transforms
required at each Runge--Kutta stage.

Our solver agrees with the Lee--Reynolds and Blaisdell--Shariff benchmark results.
We then examined three additional mean deformation cases with general
fully populated mean velocity gradient tensors, $\mathbf A(t)$.
In the General case, the grid comparison at the lower Reynolds number showed close
agreement in $K$ and $\epsilon_{ij}$. At both Reynolds numbers, remeshing
changed $K$, $\langle u_2u_3\rangle$, $\epsilon_{23}$, and
$\mathcal R^{(s)}_{23}$ by less than $0.1\%$ at the endpoint, consistent with
the smaller loss of physical resolution. In the General Shear case, however, physical resolution deteriorated
much more strongly without remeshing. Although the bulk statistics
\rev{(kinetic energy and total dissipation)} remained close,
\rev{direction-dependent turbulence statistics (such as shear stress and
the dissipation tensor component $\epsilon_{23}$)} differed substantially,
making the need for remeshing clearer.
In the Rapid General Shear case, bulk and \rev{direction-dependent turbulence
statistics} agreed closely with viscous RDT in the rapid distortion limit,
indicating a small net effect of nonlinear interactions
on these quantities, while the effect of remeshing on these statistics
remained limited.

All in all, the benchmark agreement and additional case comparisons
demonstrate the framework's ability to treat general spatially uniform mean
velocity gradients within a single formulation. These results provide a route
for generating DNS reference data over a broader range of flows
for fundamental studies of homogeneous turbulence and
the development and assessment of anisotropic turbulence closures.

\section*{Acknowledgement}
This study was supported by the Scientific and Technological Research Council of Türkiye (T\"UB\.{I}TAK), Project No. 124M535, and by the EU co-funded partnership CONNECT-NM under Grant Agreement No. 101165375. Tamerlan Hajiyev thanks {\"O}zye{\u{g}}in University for providing his scholarship.

\section*{Credit authorship contribution statement}
\textbf{Tamerlan Hajiyev:} Writing – original draft, Visualization, Validation, Software, Methodology, Investigation, Formal analysis, Data curation, Conceptualization; \textbf{Emre Sefer:} Supervision, Software, Methodology, Writing – review \& editing;  \textbf{Ivan Otić:} Writing – review \& editing, Methodology, Formal analysis; \textbf{Ismail Ari:} Writing – review \& editing, Methodology, Conceptualization; \textbf{Özgür Ertunç:} Writing – review \& editing, Supervision, Investigation, Project administration, Methodology, Funding acquisition, Formal analysis, Data curation, Conceptualization.

\section*{Declaration of competing interest}
We have no conflicts of interest to declare.

\section*{Data availability}

Data will be made available upon request.

\bibliographystyle{elsarticle-harv}
\bibliography{refs}

@techreport{rogallo1981,
  author = {Rogallo, Robert S.},
  title = {Numerical experiments in homogeneous turbulence},
  institution = {NASA Ames Research Center},
  number = {NASA TM-81315},
  year = {1981},
  url = {https://ntrs.nasa.gov/citations/19810022965}
}

@techreport{ertunc2025transport,
  author = {Ertun{\c{c}}, {\"O}zg{\"u}r},
  title = {Derivations of the transport equations for the {Reynolds} stresses, the turbulence kinetic energy and its dissipation rate},
  institution = {Fluid Dynamics and Spray Laboratory, {\"O}zye{\u{g}}in University},
  type = {ADSL Open Report},
  number = {001v1},
  year = {2025},
  month = jul,
  doi = {10.13140/RG.2.2.16536.30723}
}

@techreport{leeReynolds1985,
  author = {Lee, M. J. and Reynolds, W. C.},
  title = {Numerical experiments on the structure of homogeneous turbulence},
  institution = {Stanford University, Department of Mechanical Engineering},
  number = {TF-24},
  year = {1985},
  month = nov,
  note = {Prepared under NASA Grant NCC-2-15},
  url = {https://web.stanford.edu/group/tfsa/TF_reports/TF-024_Lee_M.J.pdf}
}

@article{genceMathieu1979,
  author = {Gence, J. N. and Mathieu, J.},
  title = {On the application of successive plane strains to grid-generated turbulence},
  journal = {Journal of Fluid Mechanics},
  volume = {93},
  pages = {501--513},
  year = {1979},
  doi = {10.1017/S0022112079002627}
}

@article{townsend1954,
  author = {Townsend, A. A.},
  title = {The uniform distortion of homogeneous turbulence},
  journal = {The Quarterly Journal of Mechanics and Applied Mathematics},
  volume = {7},
  number = {1},
  pages = {104--127},
  year = {1954},
  doi = {10.1093/qjmam/7.1.104}
}

@phdthesis{ertunc2007,
  author = {Ertun{\c{c}}, {\"O}zg{\"u}r},
  title = {Experimental and numerical investigations of axisymmetric turbulence},
  school = {Universit{\"a}t Erlangen-N{\"u}rnberg},
  address = {Erlangen, Germany},
  year = {2007}
}

@inproceedings{ertuncEtAl2007,
  author = {Ertun{\c{c}}, {\"O}zg{\"u}r and K{\"o}ksoy, {\c{C}}a{\u{g}}atay and Ray, Subhashis and Durst, Franz},
  title = {Rapid axisymmetric contraction of grid-generated turbulence and the {Reynolds}-stress models at rapid distortion limit},
  booktitle = {Proceedings of the Fifth International Symposium on Turbulence and Shear Flow Phenomena},
  address = {Munich, Germany},
  pages = {189--194},
  year = {2007},
  doi = {10.1615/TSFP5.280}
}

@article{banerjeeEtAl2009PressureStrain,
  author = {Banerjee, S. and Ertun{\c{c}}, {\"O}. and K{\"o}ksoy, {\c{C}}. and Durst, F.},
  title = {Pressure strain rate modeling of homogeneous axisymmetric turbulence},
  journal = {Journal of Turbulence},
  volume = {10},
  pages = {N29},
  year = {2009},
  doi = {10.1080/14685240903127301}
}

@article{jovanovic2003,
  author = {Jovanovi{\'c}, J. and Oti{\'c}, I. and Bradshaw, P.},
  title = {On the anisotropy of axisymmetric strained turbulence in the dissipation range},
  journal = {Journal of Fluids Engineering},
  volume = {125},
  number = {3},
  pages = {401--413},
  year = {2003},
  doi = {10.1115/1.1568355}
}

@article{yuGirimaji2006,
  author = {Yu, Dazhi and Girimaji, Sharath S.},
  title = {Direct numerical simulations of homogeneous turbulence subject to periodic shear},
  journal = {Journal of Fluid Mechanics},
  volume = {566},
  pages = {117--151},
  year = {2006},
  doi = {10.1017/S0022112006001832}
}

@article{gualtieriMeneveau2010,
  author = {Gualtieri, P. and Meneveau, C.},
  title = {Direct numerical simulations of turbulence subjected to a straining and destraining cycle},
  journal = {Physics of Fluids},
  volume = {22},
  pages = {065104},
  year = {2010},
  doi = {10.1063/1.3453709}
}

@article{clayYeung2016,
  author = {Clay, M. P. and Yeung, P. K.},
  title = {A numerical study of turbulence under temporally evolving axisymmetric contraction and subsequent relaxation},
  journal = {Journal of Fluid Mechanics},
  volume = {805},
  pages = {460--493},
  year = {2016},
  doi = {10.1017/jfm.2016.566}
}

@article{brucker2007,
  author = {Brucker, Kyle A. and Isaza, Juan C. and Vaithianathan, T. and Collins, Lance R.},
  title = {Efficient algorithm for simulating homogeneous turbulent shear flow without remeshing},
  journal = {Journal of Computational Physics},
  volume = {225},
  number = {1},
  pages = {20--32},
  year = {2007},
  doi = {10.1016/j.jcp.2006.10.018}
}

@phdthesis{zhu2019thesis,
  author = {Zhu, Ying},
  title = {Modelling and calculation for shear-driven rotating turbulence, with multiscale and directional approach},
  school = {Universit\'{e} de Lyon, \'{E}cole Centrale de Lyon},
  year = {2019},
  url = {https://bibliotheque.ec-lyon.fr/documents/TH_T2664_yzhu.pdf}
}

@article{dobson2014,
  author = {Dobson, Matthew},
  title = {Periodic boundary conditions for long-time nonequilibrium molecular dynamics simulations of incompressible flows},
  journal = {Journal of Chemical Physics},
  volume = {141},
  pages = {184103},
  year = {2014},
  doi = {10.1063/1.4901276}
}

@article{girimajiPope1990,
  author = {Girimaji, S. S. and Pope, S. B.},
  title = {Material-element deformation in isotropic turbulence},
  journal = {Journal of Fluid Mechanics},
  volume = {220},
  pages = {427--458},
  year = {1990},
  doi = {10.1017/S0022112090003330}
}

@article{iserles2000,
  author = {Iserles, Arieh and Munthe-Kaas, Hans Z. and N{\o}rsett, Syvert P. and Zanna, Antonella},
  title = {Lie-group methods},
  journal = {Acta Numerica},
  volume = {9},
  pages = {215--365},
  year = {2000},
  doi = {10.1017/S0962492900002154}
}

@article{yuGirimaji2007,
  author = {Yu, Huidan and Girimaji, Sharath S.},
  title = {Extension of compressible ideal-gas rapid distortion theory to general mean velocity gradients},
  journal = {Physics of Fluids},
  volume = {19},
  pages = {041702},
  year = {2007},
  doi = {10.1063/1.2718912}
}

@misc{dongHaeringMoser2025,
  author = {Dong, Hanqi and Haering, Sigfried and Moser, Robert D.},
  title = {A {DNS} Forcing Scheme Representing Anisotropic Production and Deformation in Homogeneous Turbulence},
  year = {2025},
  note = {78th Annual Meeting of the APS Division of Fluid Dynamics, Abstract E21.18},
  url = {https://meetings-archive.aps.org/dfd/2025/e21/18/}
}

@misc{hajiyev2026closure,
  author = {Hajiyev, Tamerlan and Sefer, Emre and Otic, Ivan and Ar{\i}, Ismail and Ertun{\c{c}}, {\"O}zg{\"u}r},
  title = {Construction of a Deployable Closure Model for the Transport of Turbulence Dissipation Rate with Hard-Constrained Physics-Informed Machine Learning},
  year = {2026},
  doi = {10.48550/arXiv.2611.23539}
}

@article{blanesMoan2006,
  author = {Blanes, Sergio and Moan, Per Christian},
  title = {Fourth- and sixth-order commutator-free {Magnus} integrators for linear and non-linear dynamical systems},
  journal = {Applied Numerical Mathematics},
  volume = {56},
  number = {12},
  pages = {1519--1537},
  year = {2006},
  doi = {10.1016/j.apnum.2005.11.004}
}

@article{pattersonOrszag1971,
  author = {Patterson, Jr., G. S. and Orszag, Steven A.},
  title = {Spectral Calculations of Isotropic Turbulence: Efficient Removal of Aliasing Interactions},
  journal = {Physics of Fluids},
  volume = {14},
  number = {11},
  pages = {2538},
  year = {1971},
  doi = {10.1063/1.1693365}
}

@article{batchelorProudman1954,
  author = {Batchelor, G. K. and Proudman, Ian},
  title = {The effect of rapid distortion of a fluid in turbulent motion},
  journal = {The Quarterly Journal of Mechanics and Applied Mathematics},
  volume = {7},
  number = {1},
  pages = {83--103},
  year = {1954},
  doi = {10.1093/qjmam/7.1.83}
}

@book{pope2000,
  author = {Pope, Stephen B.},
  title = {Turbulent Flows},
  publisher = {Cambridge University Press},
  year = {2000},
  url = {https://www.cambridge.org/core/books/turbulent-flows/69322053C06F73F7EB7124915F9256BD}
}

@article{kassinosReynoldsRogers2001,
  author = {Kassinos, S. C. and Reynolds, W. C. and Rogers, M. M.},
  title = {One-point turbulence structure tensors},
  journal = {Journal of Fluid Mechanics},
  volume = {428},
  pages = {213--248},
  year = {2001},
  doi = {10.1017/S0022112000002615}
}

@article{duraisamy2019,
  author = {Duraisamy, Karthik and Iaccarino, Gianluca and Xiao, Heng},
  title = {Turbulence Modeling in the Age of Data},
  journal = {Annual Review of Fluid Mechanics},
  volume = {51},
  pages = {357--377},
  year = {2019},
  doi = {10.1146/annurev-fluid-010518-040547}
}

@article{jangDeBruynKops2007,
  author = {Jang, Y. and {de Bruyn Kops}, S. M.},
  title = {Pseudo-spectral numerical simulation of miscible fluids with a high density ratio},
  journal = {Computers \& Fluids},
  volume = {36},
  number = {2},
  pages = {238--247},
  year = {2007},
  doi = {10.1016/j.compfluid.2005.11.003}
}

@techreport{hajiyev2026rogallo,
  author = {Hajiyev, Tamerlan and Sefer, Emre and Otic, Ivan and Ar{\i}, Ismail and Ertun{\c{c}}, {\"O}zg{\"u}r},
  title = {Direct Numerical Simulation of Incompressible Homogeneous Turbulence under General Spatially Uniform Mean Velocity Gradients: Mathematical Derivations},
  institution = {\"Ozye\u{g}in University},
  type = {ADSL Open Report},
  number = {003v1},
  year = {2026},
  month = nov,
  doi = {10.13140/RG.2.2.12057.86881}
}

@inproceedings{blaisdellShariff1996,
  author = {Blaisdell, G. A. and Shariff, K.},
  title = {Simulation and modeling of the elliptic streamline flow},
  booktitle = {Proceedings of the 1996 Center for Turbulence Research Summer Program},
  organization = {Center for Turbulence Research, Stanford University and NASA Ames Research Center},
  pages = {433--446},
  year = {1996},
  url = {https://ntrs.nasa.gov/citations/19970014678}
}

@inproceedings{banerjeeErtuncDurst2008,
  author = {Banerjee, Sankha and Ertun{\c{c}}, {\"O}zg{\"u}r and Durst, Franz},
  title = {Anisotropy properties of turbulence},
  booktitle = {Proceedings of the 13th {WSEAS} International Conference on Applied Mathematics},
  pages = {26--57},
  year = {2008}
}

@article{zusiPerot2013,
  author  = {Zusi, Christopher J. and Perot, J. Blair},
  title   = {Simulation and modeling of turbulence subjected to a period of uniform plane strain},
  journal = {Physics of Fluids},
  volume  = {25},
  number  = {11},
  pages   = {110819},
  year    = {2013},
  doi     = {10.1063/1.4821450}
}

@article{mansour1994,
  author  = {Mansour, Nagi N. and Wray, Alan A.},
  title   = {Decay of isotropic turbulence at low {R}eynolds number},
  journal = {Physics of Fluids},
  volume  = {6},
  number  = {2},
  pages   = {808--814},
  year    = {1994},
  doi     = {10.1063/1.868319}
}

@incollection{mansourCambonSpeziale1991,
  author    = {Mansour, N. N. and Cambon, C. and Speziale, C. G.},
  title     = {Single point modeling of initially isotropic turbulence under uniform rotation},
  booktitle = {Annual Research Briefs 1991},
  publisher = {Center for Turbulence Research, Stanford University and NASA Ames Research Center},
  pages     = {159--167},
  year      = {1991},
  url       = {https://web.stanford.edu/group/ctr/ResBriefs/1991/15_MANSOUR.pdf}
}

\end{document}